\documentclass[12pt]{iopart}
\usepackage{graphicx}
\usepackage{color}
\usepackage{url}

\begin{document}

\title[EUV Al spectra in EAST]{Line identification of extreme ultraviolet
	spectra from aluminum ions in
	EAST Tokamak plasmas}
	
\author{Fengling Zhang$^{1,2}$, 
Ling Zhang$^{\dag 1}$, Wenming Zhang$^{1,2}$,Yunxin Cheng$^{1}$, Ailan Hu$^{1}$,  
Xiaobin Ding$^3$, Shigeru Morita$^4$, Zhengwei Li$^{1,5}$, Zhen Zhou$^{1,2}$, Yiming Cao$^{1,5}$, Jiuyang Ma$^{1,2}$, Zhehao Xu$^{1,5}$, Lang Xu$^{1,2}$, Chenxi Zhou$^{2}$, Yinxian Jie$^{1}$, and Dar\'{\i}o Mitnik$^{* 6,1}$ 
}

\address{$^1$ Institute of Plasma Physics, Hefei Institutes of Physical Science, Chinese Academy of Sciences, Hefei 230031, China}
\address{$^2$ University of Science and Technology of China, Hefei 230026, China}
\address{$^3$ Key Laboratory of Atomic and Molecular Physics and Functional Materials of Gansu Province, College of Physics and Electronic Engineering, Northwest Normal University, Lanzhou 730070, China}
\address{$^4$ National Institute for Fusion Science, Toki 509-5292, Gifu, Japan}
\address{$^5$ Anhui University, Hefei 230601, China}
\address{$^6$ Instituto de Astronom\'{\i}a y F\'{\i}sica del Espacio (CONICET - Universidad de Buenos Aires), Buenos Aires 1428, Argentina}

\ead{$^{*}$dmitnik@df.uba.ar, $^{\dag}$zhangling@ipp.ac.cn}

\vspace{10pt}
\begin{indented}
\item[]August 2023 
\end{indented}

\begin{abstract}
	
Extreme ultraviolet (EUV) spectra emitted from aluminum in the 5–340 \AA~wavelength range were observed in Experimental Advanced Superconducting Tokamak (EAST) discharges. Several spectral lines from aluminum ions with different degrees of ionization were successfully observed with sufficient spectral intensities and resolutions using three fast-time-response EUV spectrometers.
The line identification uses three independent state-of-art computational codes for the atomic structure calculations, which provide the wavelengths and the collisional and radiative transition rate coefficients.
These programs are {\sc hullac} (Hebrew University - Lawrence Livermore Atomic Code),  {\sc autostructure}, and {\sc fac} (Flexible Atomic Code). 
Using three different codes allows us to resolve some ambiguities in identifying certain spectral lines and assess the validity of the theoretical predictions.
	
\end{abstract}
\vspace{2pc}
\noindent{\it Keywords}: {plasma spectroscopy, Tokamak spectra, relativistic atomic structure}
\submitto{\PS}


\section{INTRODUCTION}

In magnetic confinement fusion Tokamak devices, there are usually various impurities in addition to the working particles. 
The primary sources of them are due to the interaction between the edge plasma and plasma-facing components (PFCs), like sputtering \cite{Langley84,Eckstein91} and desorption \cite{Zhao04} processes.
Impurities can penetrate the central plasma causing significant power losses, diluting the fuel ions, and disrupting the steady-state operation of the plasma. 
Therefore, spectroscopic measurements of line emissions from impurity ions are crucial to monitor their behavior and minimize their effects on plasma performance.

The Experimental Advanced Superconducting Tokamak (EAST) is an advanced steady-state plasma device described in several reports (see, for example, \cite{Wu07,Wan17,Wan19,Wan22}).
Various PFCs are located in different regions of EAST, protecting the vacuum vessel, the heating systems, and the diagnostic components from the plasma particles and heat loads \cite{Song10}.
Examples of PFC are the tungsten divertors \cite{Song10,Wan22}, the first wall made of molybdenum bricks \cite{Song10}, the tungsten blocks assembly with CuCrZr employed in the guard limiter for the lower hybrid waveguide antennas \cite{Liu17,ZhangLiangliang18}, the carbon bricks where the neutral beam injection (NBI) is deposited \cite{Chundong15}, and other physical diagnostic systems installed on the vacuum chamber wall. 

In our previous works, we have observed and identified the EUV spectra of low-Z impurities \cite{Lei21} such as helium, lithium, boron, carbon, oxygen, neon, silicon, argon, and high-Z impurities  \cite{ZhangLing19,ZhangWenmin22} such as iron, copper, molybdenum, and tungsten. 
We used two sets of fast-time-response EUV spectrometers \cite{ZhangLing15,Xu21} called `EUV\_Short' and `EUV\_Long' 
working at 8-130  \AA~ and 20-500 \AA. 
In that way, we have established a primary database of impurity EUV spectra in EAST plasma. 
However, in these spectra, we still found some unrecognized lines that did not come from the impurity elements that have been identified. 
During the 2021-2022 EAST experimental campaigns, a novel electromagnetic probe 
array (EMPA) diagnostic has been adopted \cite{Yan14,Meng22,Lan23}.
One fundamental measurement unit of this EMPA is the carefully designed 3D magnetic probe, which can measure the magnetic fluctuations in poloidal, radial, and toroidal directions simultaneously. 
Each bobbin in this probe is made from high-temperature-resistant 95 alumina
ceramics (Al$_2$O$_3$).
In the present work, by using several high-intensity spectral lines of intrinsic low-Z impurities for calibration, through careful comparison and analysis with relevant literature and with the theoretical spectra obtained using three independent state-of-the-art computational codes, it is possible to confirm with certainty that such unrecognized spectral lines originate in the radiation 
emission of Al ions.

As far as we know, this is the first time aluminum ions have been identified as an inherent impurity in Tokamak plasmas.
Valero and Goorvitch reported spectral lines of Al VI to Al X ions in the range of 48 to 335 \AA~in laser-produced plasma experiments \cite{Valero72}. 
Other laser-produced plasma experiments have also observed resonant transition lines of Al hydrogen-like  \cite{Boiko78,Shlyaptsev16} and helium-like \cite{Boiko77,Couturaud77,Channprit17} aluminum ions.
Gu et al. \cite{Gu11} observed aluminum ions's L-shell emission spectra in the 40-170 \AA~range in the Livermore electron beam ion trap (EBIT) device.  They identified the lines by using a detailed collisional-radiative (CR) model.
In addition, the EUV spectra of highly ionized Al ions have been observed during a laser blow-off experiment using an aluminum target for an NBI-heated plasma created in the TJ-II stellarator \cite{McCarthy16}. However, they only observed lines with wavelengths higher than 175 \AA.

The paper is organized as follows. 
In Section~\ref{sec:experimental}, we briefly describe the spectrometers employed and some spectra observed in the EAST experiments, in which Al lines can be identified.
Section~\ref{sec:theory} describes the three different atomic structure computational codes used in our calculations. 
The atomic data these codes provide is processed in a collisional-radiative model, also described in this section.
That allows us to identify the aluminum spectral lines with certainty, as detailed in Section~\ref{sec:spectra}.
Finally, Section~\ref{sec:summary} summarizes our findings.

\section{Experimental setup}
\label{sec:experimental}

\begin{center}
\begin{figure}
\linespread{0.6}
\centering\includegraphics[width=0.5\textwidth]
	{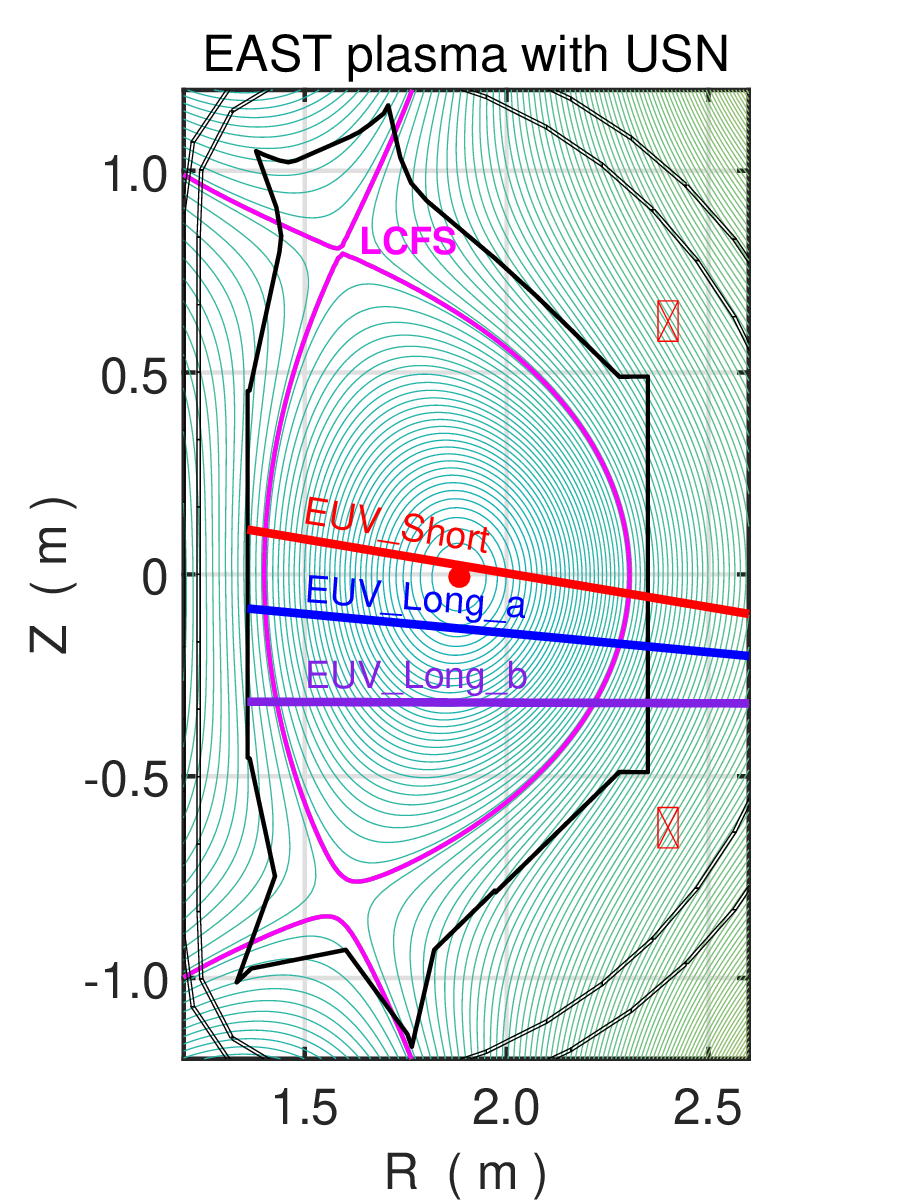}   
	\caption{LOS of the two fast-time-response EUV spectrometers.
		Red line: EUV\_Short. Blue line: EUV\_Long\_a.
		Purple line: EUV\_Long\_b.}
	\label{fig:LOS}
\end{figure}   
\end{center}

During the EAST upgrade in 2021 \cite{Song22}, two new fast EUV spectrometers were supplemented the previous set, making up a total of four fast EUV spectrometers, namely EUV\_Short, EUV\_Long\_a, EUV\_Long\_b, and EUV\_Long\_c. 
All the spectrometers are equipped with a laminar-type varied-line-spacing concave holographic grating (1200 and 2400 grooves/mm for long and short, respectively). 
They are fixed at grazing incidence ($87.0^\circ$ and $88.6^\circ$ for long and short, respectively) with a narrow entrance slit width of 30 $\mu$m to increase the spectral resolution. 
A back-illuminated charge-coupled device (CCD) with a sensitive area of 26.6 $\times 6.6$ mm$^2$ 
($1024 \times 255$ pixels, $26 \times 26~\mu$m$^2$/pixel) is used for recording the spectral image. 
The spectra are recorded every 5 ms when the CCD is operated with full vertical binning mode. 
Previously, the two spectrometers needed to adjust the stepper motor to control the CCD's position and change the wavelength's observation range.
Now, the four spectrometers have achieved full coverage in the 
5-500 \AA~band without changing the CCD position. 
In the present experiment, the EUV\_Long\_c spectrometer was not operating; therefore, we analyzed only the 5-380 \AA~band. 
The wavelength calibration is performed by cubic polynomial fitting with many well-known spectral lines covering the observable range \cite{Lei21}. 
The line of sight (LOS) for the other three spectrometers is indicated  with the EAST plasma cross-section in Fig.~\ref{fig:LOS}. 

The uncertainties and resolutions of the different spectrometers 
have been carefully studied in Refs. \cite{ZhangLing15} and \cite{Xu21}. 
The EUV\_Short spectrometer has a high spectral resolution of 
$\Delta \lambda_0=$4–6 pixels ( $\Delta \lambda_0$(FWHM)=0.1 \AA~at 20 \AA, 
and 0.17 \AA~at 50 \AA), where $\Delta \lambda_0$ is defined 
as the full width at the foot position of a spectral line, 
and $\Delta \lambda_0$(FWHM) is the standard definition of 
the spectral resolution, full width at half maximum.
For the EUV\_Long spectrometer, it is found that the
spectral resolution is slightly poorer at 
shorter wavelengths between 20-50 \AA, e.g., 
$\Delta \lambda_0$(FWHM)=0.8 \AA~and 0.5 \AA~at 30 \AA~and 50 \AA, 
respectively, while the spectral resolution is 
excellent at the 80-400 \AA~wavelength range, e.g., 
$\Delta \lambda_0$ is between 3 and 5 pixels 
($\Delta \lambda_0$(FWHM)=0.22 at 100 \AA~and 0.30 \AA~at 200 \AA,
respectively).
The uncertainty of the wavelength at each pixel for different
observation wavelength intervals introduced by different 
calibration methods is $|\Delta \lambda|<0.03$ \AA~ for EUV\_Short, 
and $|\Delta \lambda|<0.08$ \AA~ for EUV\_Long.
For the EUV\_Short spectrometer, the reciprocal linear dispersion 
$\frac{d\lambda}{dX}$ varies
from 1.3 \AA/mm to 4.14 \AA/mm (0.03–0.1 \AA/pixel) when the wavelength 
increases from 10 \AA~to 130 \AA.
For EUV\_Long, the reciprocal linear dispersion 
varies from 3.05 to 10.78 \AA/mm
(0.08–0.28 \AA/pixel) in the wavelength range of 20-500 \AA.

Aluminum spectra are observed with a singular plasma sputtering event, 
which is abnormal and rare compared to Fe and Cu sputtering 
(Fe widely existed in the first wall, and Cu is the antenna material). 
Due to the transport process in the plasma after sputtering,
the ionization stages of Al ions and their distribution change quickly 
with time; some emission lines only existed for unique or very few 
continuous frames. Therefore, spectra presented in this work 
are just from one single frame without accumulation, 
which means more considerable statistical uncertainties should be 
considered for these lines. 
Since statistical accumulation is unavailable, the
resolution plays a role in wavelength determination. 
Moreover, in this work, only the maximum peak
position at a particular pixel is used to determine the wavelength 
of the Al emission lines.  
Therefore, for a safe estimation, we can consider one-pixel error 
(the reciprocal linear dispersion $\frac{d\lambda}{dx}(\lambda)$, 
see \cite{Lei21}) as an additional factor to consider in the total error 
evaluation.
For the particular time frame in which the Al lines appear, 
each spectrometer has only one fixed initial CCD position ($X_0$), 
and each one covers a different range: EUV\_Short is used for 
5-50 \AA,  EUV\_Long\_a for 50-180 \AA~and EUV\_Long\_b for $\lambda > 180$ \AA. 
Assuming the considered safe condition for the uncertainties
and a 1-pixel error in the lecture, we can estimate an overall error
of $\Delta \lambda < 0.1$ \AA~for EUV\_Short ($\lambda < 50$ \AA), and 
$\Delta \lambda < 0.2-0.3$ \AA~for EUV\_Long, 
for the range  $50 < \lambda < 300$ \AA. 

\begin{figure}[h!]
\centering\includegraphics[width=0.95\textwidth]
	{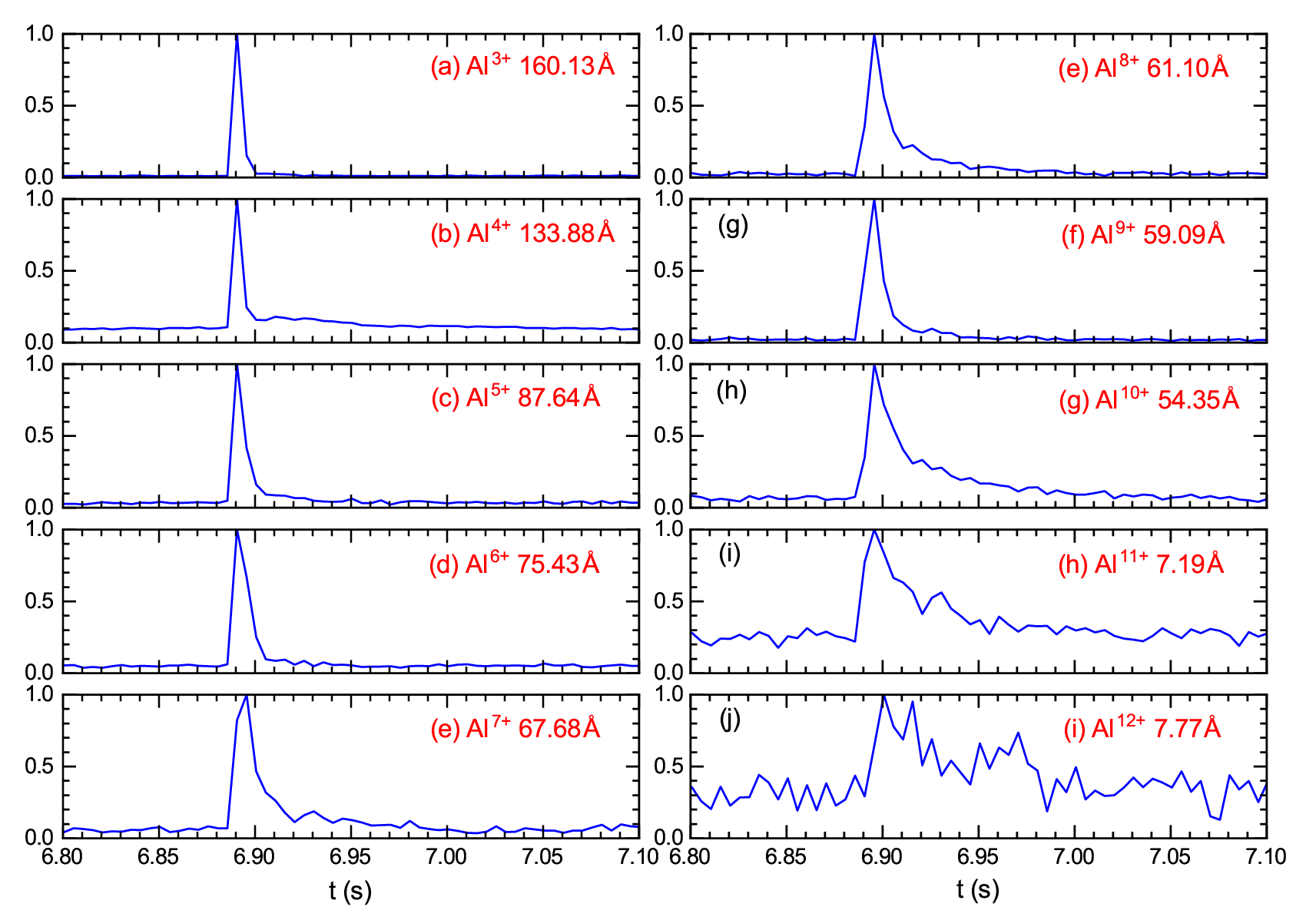}    
	\caption{Time evolution of the Al$^{3+}$ to Al$^{12+}$ dominant lines emission during the EAST 108709 discharge.}
	\label{fig:shot}
\end{figure}

Once we establish that unknown spectral lines can come from the Al presence, the next step is to look at the temporal behavior of some dominant lines of their different ions. 
An illustrative example is in Fig.~\ref{fig:shot}, where the Al lines arise in a spontaneous sputtering event during the EAST 108709 discharge. 
A sudden increase in the intensities of the Al spectral lines occurred at 6.885 seconds, reached a maximum, and then returned to the average value after 15 ms. The precise time of the maximum intensities depends on the ionization degree, being $t=6.890$ s for the lower charges (Al$^{3+}$ to Al$^{5+}$) and $t=6.895$ for the other ions, except Al$^{12+}$, in which the maximum intensity arises at $t=6.900$ s.

\begin{figure}[h!]
	\centering\includegraphics[width=0.95\textwidth]
	{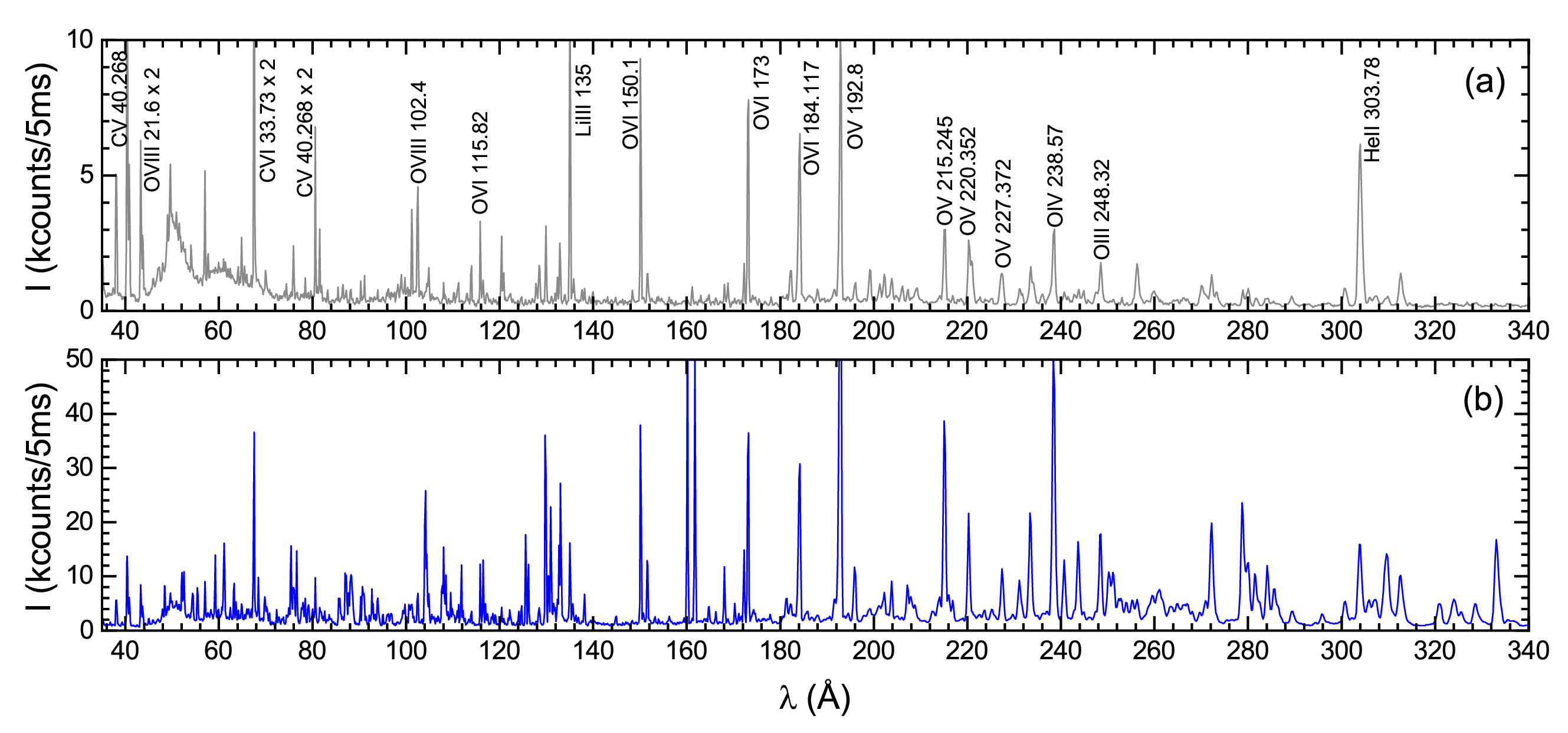}   
	\caption{Full spectra of the EAST 108709 discharge taken (a) at time $t<6.890$ s, before the Al burst, and (b) at $t=6.895$ s, when the Al intensity reaches the maximum value.}
	\label{fig:spectrum2}
\end{figure}

The emission lines corresponding to the Al ions can be easily found by the comparison of the spectra obtained before ($t<6.890$ s) and during ($t=6.895$ s) the Al burst, as is shown in Figure~\ref{fig:spectrum2}. 
In this particular discharge, the electron temperature of the plasma core was $T_e(0)=2.5$ keV, and the chord-average electron density $n_e=2.6 \times 10^{19}$ m$^{-3}$. Both parameters have been provided by a Thomson scattering system \cite{Han18} and reflectometer measurements \cite{Wang19}. 

\begin{figure}[h!]
	\centering\includegraphics[width=0.95\textwidth]
	{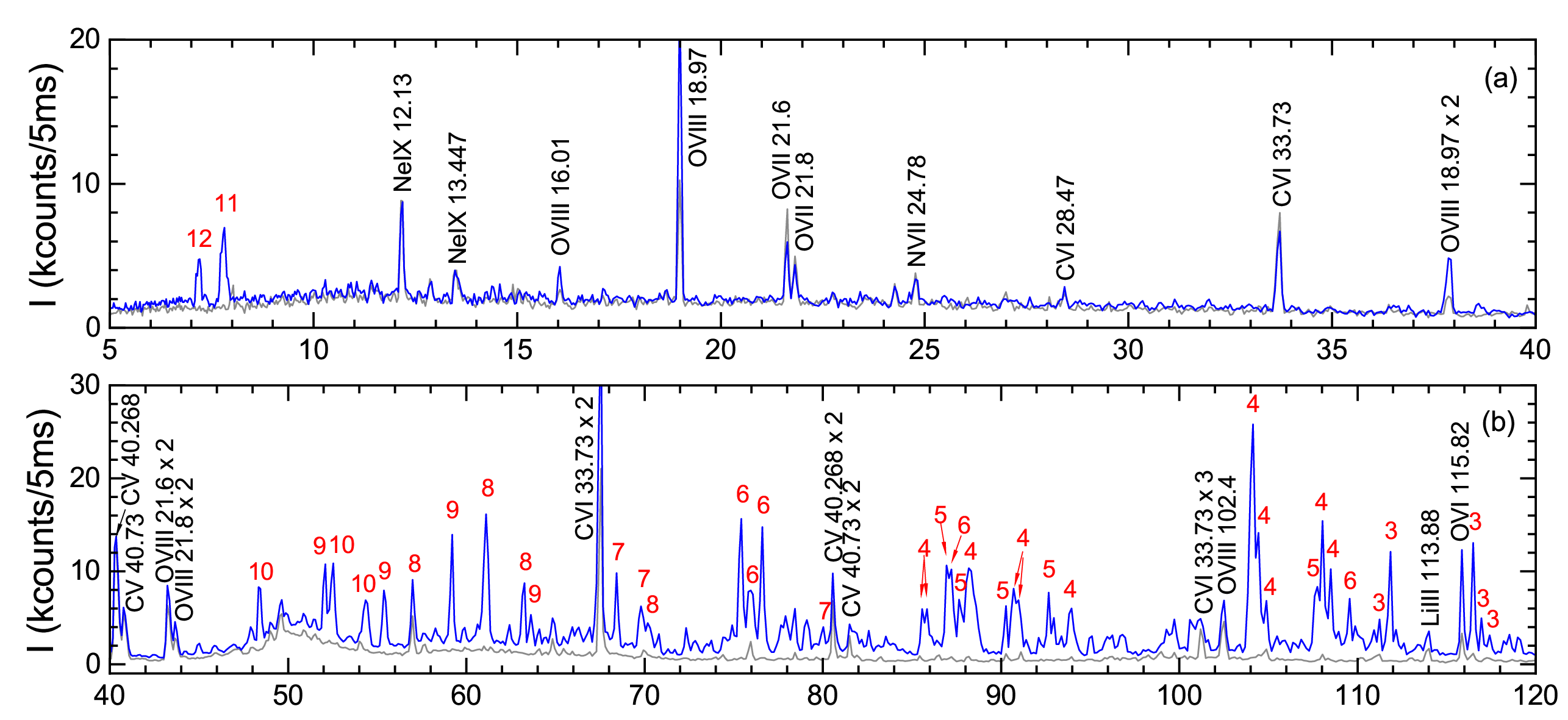}   \\
	\centering\includegraphics[width=0.95\textwidth]
	{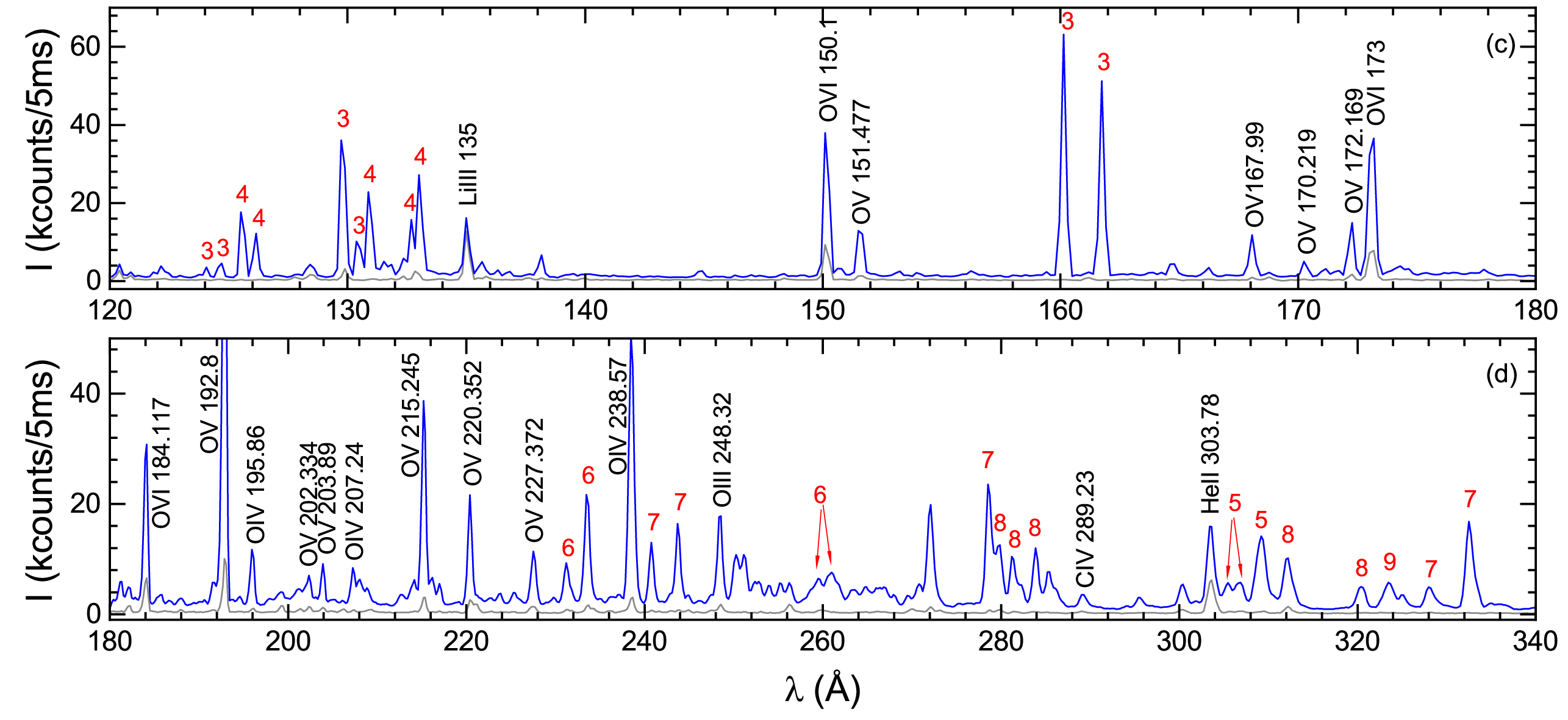}   
	\caption{EUV spectra observed in the EAST experiment and lines identified as belonging to aluminum ions 
	(red labels).
	The numbers $n$ means that the corresponding ion is Al$^{n+}$.
	Gray curve: spectrum before the Al burst. Blue curve: spectra at maximum Al emission.}
	\label{fig:fullAl}
\end{figure}

Figure~\ref{fig:fullAl} shows the aluminum spectra observed in the 5-340  \AA~ band.
In this, we indicate the stronger lines used in the calibration method \cite{Lei21,ZhangWenmin22}. Most of them belong to Li, C, and especially O ions. As is shown in the Figure, the highly ionized Al (H- and He-like) only contribute to the spectra at short wavelengths, and most of the Al lines appear between 50 \AA~and 135 \AA.  The dominant lines around 160 \AA~belong to Al$^{3+}$. We are not plotting the spectra beyond 340 \AA~since the intensities are very small. In Section \ref{sec:spectra}, we will provide a detailed analysis of all the lines that we have been able to identify with certainty.

\clearpage
\section{Theoretical methods}
\label{sec:theory}

The spectral lines have been identified with three separate independent calculations. 
We used the atomic structure computational 
codes {\sc hullac} (Hebrew University - Lawrence Livermore Atomic Code) \cite{Klapisch77,BarShalom01,Klapisch77}, the {\sc as} (AutoStructure) code \cite{Badnell86,Badnell97,Badnell11}, and the {\sc fac} (Flexible Atomic Code) \cite{Gu08}.

Since the three codes are very well-known in astrophysics and fusion communities, we only briefly summarize their main features. {\sc hullac} is a fully relativistic atomic structure package of codes that provides the transition energies and the corresponding rate coefficients, among other quantities. 
The code implements the relativistic parameter potential method to solve the Dirac Hamiltonian. 
Full multiconfiguration wavefunctions are used to compute the radiative transition rates; thus, configuration mixing, and therefore, correlation effects, are included in the calculations.
The Breit interaction and quantum electrodynamic  (vacuum polarization and the self-energy) corrections are treated as second-order perturbations.
The {\sc autostructure} code allows calculating energy levels, oscillator strengths, excitation, photoionization cross sections, and autoionization rates, among other quantities. 
These can be calculated with configuration resolution (configuration average CA), term resolution (LS coupling),
or level resolution (intermediate coupling IC) using semi-relativistic kappa-averaged wavefunctions. It also allows the inclusion of mixing configurations. 
{\sc fac} employs a fully relativistic approach, solving the Dirac equation. 
Quantum-electrodynamic effects, mainly arising from Breit interaction,
vacuum polarization, and electron self-energy are all included with
standard procedures in the code. 
Like the other programs mentioned, {\sc fac} has been widely employed to interpret laboratory and astrophysical spectroscopic data.

\begin{center}
\begin{figure}
\linespread{0.6}
\centering\includegraphics[width=0.95\textwidth]
		{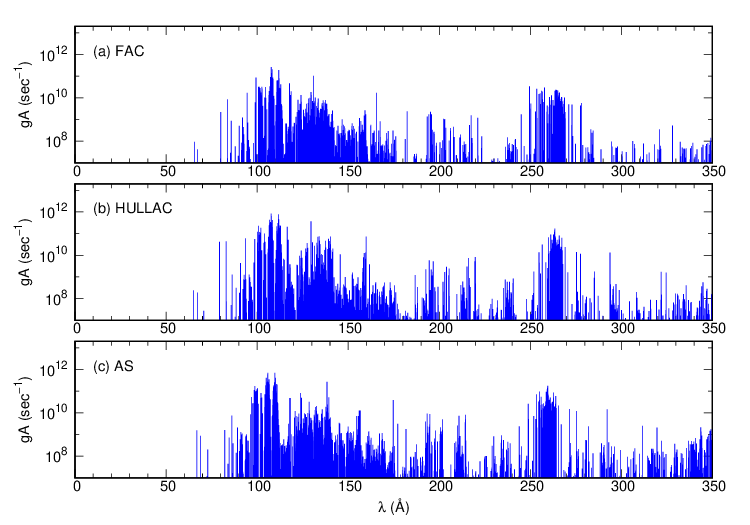}  
\caption{Comparisons of radiative transition rate coefficients  
	  for Al$^{3+}$ calculated by (a) {\sc fac}, (b) {\sc hullac}, and (c) {\sc as}.}
\label{fig:compA}
\end{figure}                      
\end{center}

The present work allows us to compare the results obtained with these three codes and evaluate their accuracy. The general agreement between them is very good; roughly, the wavelengths differ by a factor close to 2  \AA. Although the codes allow us to adjust the energies with experimental values, we prefer to show the comparisons in an unrefined form, to get a glimpse of the reliability of the basic calculations.
As an example, we show, in Fig.~\ref{fig:compA}, more than 16000 transition lines corresponding to the Al$^{3+}$ ion. The Figure shows only the radiative transition rate coefficients. 

Identifying spectral lines by counting only the radiative transition probabilities in such complex cases is impossible.
A further detailed CR model is needed to compare with the experimental spectra. 
For this task, we assume that the dominant populations of impurities in the plasma are those of the ground and metastable states of the various ions, and only these states are included in the fractional ion abundance calculations. 
The population of the excited states of each ion is obtained afterward, assuming that these levels are in equilibrium with the corresponding 
metastable states through collisional and radiative processes. 
This approximation allows us to decouple the level populations of each ion partially.
To determine the population $n_j$ of the excited levels $j$ of each ion, we solve, for a given electronic density and temperature, the collisional-radiative equations:
\begin{eqnarray}
\frac{d n_j}{dt} &=& 
\sum_{k>j} n_k \left(A_{kj} + n_e \, Q_{kj} \right) + 
\sum_{i<j} n_i \, n_e \, Q_{ij}  - 
\nonumber \\
& & - 
n_j \left(  \sum_{i<j}    \left(A_{ji} + n_e \, Q_{ji} \right) +  
\sum_{k>j} n_e \, Q_{jk}  \right) \,\, ,
\label{eq:crm} 
\end{eqnarray}
where $n_e$ is the electronic density, 
$A_{ji}$ are the radiative rate coefficient for transitions 
from level $j$ to level $i$.  
$Q_{ji}$ are the electron-impact excitation rate coefficients if $i>j$, 
and the deexcitation rates if $i<j$.
All these rate coefficients have been calculated with the computational codes described above.

The strength of a spectral line produced by the radiative transition 
$j \rightarrow i$ is proportional to its emissivity
\begin{equation}
\epsilon_{ji} = n_j \, A_{ji}  \, .
\end{equation}
In order to produce the full synthetic spectra (composed of all the different ions), it is necessary to scale the individual strengths of each ion with the corresponding fractional ion abundance factor $n^z$,  obtained after solving the charge-state distribution equations.
For Tokamak plasmas, one can assume a coronal equilibrium approximation, in which the population distribution over all ionization stages is described by a set of equations of the type:
\begin{eqnarray}
\frac{n^{z+1}}{n^{z} }  &=& \frac{S_z}{\alpha_{z+1}}
\label{eq:coronal} 
\end{eqnarray}
where $n^z$ is the ground state population of the ion with charge $z$, 
$S_z$ is its electron impact ionization, and $\alpha_{z+1}$ is the 
radiative recombination rate coefficient from the next $(z+1)$-charged ion. 
The ion ground population fractions are normalized so that $\sum_z n^z = 1$.
In the present work, the ionization and radiative rates have been taken from the {\sc adas} \cite{ADAS}. 
In some cases, we also used the fractional ionization abundance factors from the {\sc flychk} database \cite{FLYCHK}, although this generally remains the same identification of the lines. 
The model used can be improved. 
In particular, the coupling between neighboring ions should be explicitly considered by including ionization and radiative recombination for every excited level and not only through the respective ground states. 
We are currently facing the task of performing extended 
collisional-radiative model calculations, allowing us to assess the different approximations used here. 

\section{Spectra of Aluminum ions}
\label{sec:spectra}
\subsection{Al$^{3+}$}
 
\begin{center}
\begin{figure}[h!] 
\centering\includegraphics[width=0.95\textwidth]
		{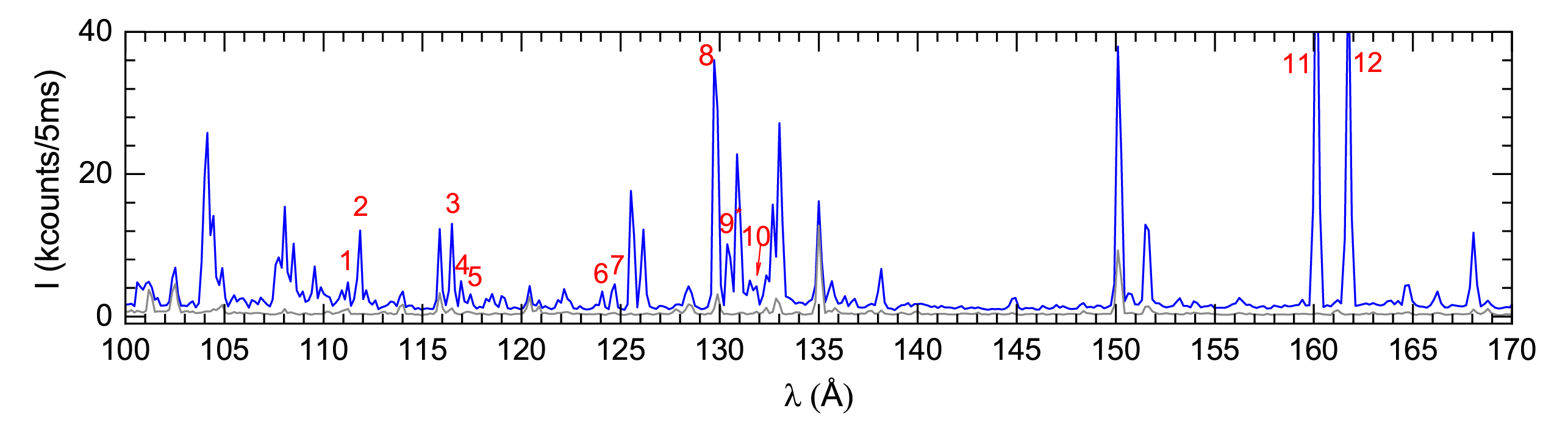}   
\caption{EAST spectrum and the lines identified as belonging to Al$^{3+}$. 
Gray curve: spectrum before the Al burst. Blue curve: spectra at maximum Al emission. 
The numbers indicate the lines (Key) described in Table~\ref{table:Al3}.}
\label{fig:Al3}
\end{figure}                      
\end{center}

The Al$^{3+}$ ion is isoelectronic to the Ne; its ground configuration is 
$1s^{2}2s^{2}2p^{6}$. 
Since it is a closed-shell ion, its ionization energy is relatively high (119.99 eV) compared to the neighboring ion stages. 
Therefore, it is supposed to have a large and broad predominance in the fractional abundances distribution (in this case, around 10 eV) over a wide range of temperatures.
Based on the spectral characteristics of Al$^{3+}$ ion described in the EBIT experiment \cite{Gu11} as well as the experimental wavelength values reported in the NIST database, we have managed to identify 12 spectral lines in the tokamak spectra, which we show in Fig.~\ref{fig:Al3}.

For the theoretical calculations, we include in our model, besides the ground configuration 
$2s^{2}2p^{6}$, the 
$2s^{2}2p^{5}nl$ ($n=3,4,5)$, $2s^{2}2p^{4}3l^{2}$, and the $2s2p^{6}nl$ ($n=3,4,5)$ configurations, including all the mixing between these configurations. 
The resultant structure has 366 levels. Due to both the existence of strong transitions and the low charge of the ion, only electron dipole-allowed transitions have been considered, resulting in more than 16000 radiative rates.
In this system, the transitions with the largest radiative rate coefficients are the 
$2s^{2}2p^{4}3d^{2} \rightarrow 2s^{2}2p^{5}3d$, with wavelengths close to 110 \AA~and Einstein coefficient values of about $10^{11}$ s$^{-1}$.
To facilitate reading in this and the following sections, we summarize these data in 
Table~\ref{table:Al3res}.

\begin{table}[h!]
\caption{Description of the theoretical model used for the calculation of 
Al$^{3+}$ spectra. $I_{\mathrm{p}}$: Ionization Energy. 
$T_{\mathrm{max}}$: Temperature of maximum abundance. 
$N$: Number of levels considered in CRM, 
$N_{E1}$: Number of dipole-allowed radiative transitions. }
\begin{indented}
\item[]\begin{tabular}{ccccc}
\br
\multicolumn{1}{c|}{Ground Configuration} & 
\multicolumn{1}{c|}{$I_{\mathrm{p}}$ (eV)} & 
\multicolumn{1}{c|}{$T_{\mathrm{max}}$ (eV)} & 
\multicolumn{1}{c|}{$N$}  &
\multicolumn{1}{c}{$N_{E1}$}    \\
\cline{1-5}
\multicolumn{1}{c|}{$1s^22s^22p^6$ (Ne-like)}      &
\multicolumn{1}{c|}{120}   & 
\multicolumn{1}{c|}{10}    & 
\multicolumn{1}{c|}{366}    &
\multicolumn{1}{c}{16208}      \\ 
\cline{1-5}  
\mr
\multicolumn{1}{l}{Even Configurations:} &
\multicolumn{3}{l}{  
$1s^{2}2s^{2}2p^{6}$,   
$1s^{2}2s^{2}2p^{5}nl$ $(n=3,4,5 ~;~ l=1,3)$, }  \\
\multicolumn{1}{l}{} &
\multicolumn{4}{l}{ 
$1s^{2}2s^{2}2p^{4}3l^{2}$ $(l=0,1,2)$, 
$1s^{2}2s2p^{6}nl$  ($n=3,4,5 ~;~ l=0,2,4)$ }   \\
\cline{1-5}
\multicolumn{1}{l}{Odd Configurations:} &
\multicolumn{4}{l}{  
$1s^{2}2s^{2}2p^{5}nl$ $(n=3,4,5 ~;~ l=0,2,4)$, }  \\
\multicolumn{1}{l}{} &
\multicolumn{4}{l}{ 
$1s^{2}2s^{2}2p^{4}3s3p$, 
$1s^{2}2s2p^{6}nl$ $(n=3,4,5 ~;~ l=1,3)$ }   \\
\br
\label{table:Al3res}
\end{tabular}
\end{indented}
\end{table}

However, our collisional radiative model shows that the upper levels of these transitions are insufficiently populated to produce a significant contribution to  the spectra. 
The transitions $2s2p^6nl \rightarrow 2s^22p^6$, with wavelengths shorter than 100 \AA~and high radiative rate coefficients, do not appear in the tokamak spectra. This is probably because these levels are above the ionization limit, allowing  autoionization channels that effectively quench the radiative path. 
All of these speculations must be validated with an extended radiative-collisional model, which exceeds the objectives of this work and which we will present in a forthcoming publication.
The next important transitions are from the odd configurations having a hole in the $2p$ shell to the ground state, i.e., $(2p^5nl)_{J=1} \rightarrow 2p^6$. 
These transitions produce the prominent lines observed in the spectra, marked in Figure~\ref{fig:Al3} and described in Table~\ref{table:Al3}. 
Some are blended with other ion's lines and cannot be identified with certainty. 
These are labeled with an asterisk, for example, line 9$^*$, which appears at the same wavelength as a 
$2p^63s \rightarrow 2p^5$ transition from Al$^{4+}$ (line 14$^*$ in Table~\ref{table:Al4}). 
The wrap-up of our observations and preparatory calculations shows that the Al$^{3+}$ spectra can be characterized by three strong transitions, the $3d \rightarrow 2p$ labeled as 8 in the Table, and the last two $3s \rightarrow 2p$ 
labeled as 10 and 11.
As is shown in ~\ref{table:Al4}), both {\sc fac} and {\sc as} calculations are in very good agreement with the experimental results, between a 2 \AA~error range. 
For some reason, {\sc fac} generally predicts wavelengths higher than the experimental values, whereas {\sc as} results are below the measurements. The results obtained with {\sc hullac} are in remarkable agreement with the experimental values, with less than a 0.3\% discrepancy between them in most cases.

\begin{table}[h!]
\caption{EAST experimental (Exp.) and theoretical lines ({\sc fac}, {\sc hullac}, and {\sc autostructure}), 
for Al$^{3+}$.
The values from other experiments and the experimental wavelengths quoted in 
NIST are also displayed.
The column `Key' indicates the transitions labeled in Fig.~\ref{fig:Al3}.
The symbol $^*$ means that the line is blended with another ion's line, 
as explained in the text. 
We use the usual relativistic nomenclature $j_{\pm} \equiv l \pm \frac{1}{2}$. 
The last subscript means the total $J$ quantum number of the level. 
Wavelengths are in \AA.}
\label{table:Al3}
\begin{indented}
\item[]\begin{tabular}{lcccccccc}
\br
Key & Exp. & {\sc fac} & {\sc hullac} & AS & Previous Exp. & NIST & Lower & Upper \\
\mr
1   & 111.24 & 112.49 & 111.36 & 109.74  & 111.301$^a$ & 111.196& $2p^6$ & 
$(2p^1_{-}2p^4_{+}5d_{-})_1$ \\  
2   & 111.85 & 112.90 & 111.77 & 110.22  & 111.711$^a$ & 111.781 & $2p^6$ & 
$(2p^2_{-}2p^3_{+}5d_{+})_1$  \\  
3	& 116.49 & 117.76 & 116.56 & 114.80  & 116.546$^a$ & 116.464& $2p^6$ & 
$(2p^1_{-}2p^4_{+}4d_{-})_1$ \\
4	& 116.96 & 118.29 & 117.08 & 115.42  & 117.070$^a$ & 116.921& $2p^6$ & 
$(2p^2_{-}2p^3_{+}4d_{+})_1$ \\  
5	& 117.42 & 118.76 & 117.55 & 115.95  &   & 117.377 & $2p^6$ & 
$(2p^2_{-}2p^3_{+}4d_{-})_1$ \\ 
6	& 124.07 & 125.49 & 124.18 & 122.39  & 124.097$^a$ & 124.030& $2p^6$ & 
$(2p^1_{-}2p^4_{+}4s)_1$ \\  
7	& 124.71 & 126.02 & 124.70 & 122.98  & 124.619$^a$ & 124.550& $2p^6$ & 
$(2p^2_{-}2p^3_{+}4s)_1$ \\  
8	& 129.73 & 130.95 & 129.57 & 127.46  & 129.793$^a$ & 129.730& $2p^6$ & 
$(2p^1_{-}2p^4_{+}3d_{-})_1$ \\  
$9^*$	& 130.39 & 131.89 & 130.40 & 128.47  & 130.473$^a$ & 130.39 & $2p^6$ & 
$(2p^2_{-}2p^3_{+}3d_{+})_1$  \\  
10	& 131.86 & 133.30 & 131.72 &  129.91  &   & 131.647 & $2p^6$ & 
$(2p^2_{-}2p^3_{+}3d_{-})_1$ \\
11	& 160.13 & 162.44 & 159.79 & 158.84  & 160.089$^a$ & 160.074& $2p^6$ & 
$(2p^1_{-}2p^4_{+}3s)_1$ \\	 
12	& 161.74 & 164.61 & 161.64 & 160.70  & 161.712$^a$ & 161.688& $2p^6$ & 
$(2p^2_{-}2p^3_{+}3s)_1$  \\
\br		 
\end{tabular}
\item[] $^a$ Livermore EBIT experimental results \cite{Gu11}.
\end{indented}
\end{table}

\clearpage
\subsection{Al$^{4+}$}

The Al$^{4+}$ ion is isoelectronic to the F, having a ground configuration $1s^22s^22p^5$.
The ionization energy is about 152 eV; however, its maximum abundance is roughly predicted to be around 20 eV. 
For the identification, we included the 740 levels listed on 
Table~\ref{table:Al4res}, and about 75000 dipole-allowed radiative transitions. 

\begin{center}
\begin{figure}[h!]
\centering\includegraphics[width=0.95\textwidth]{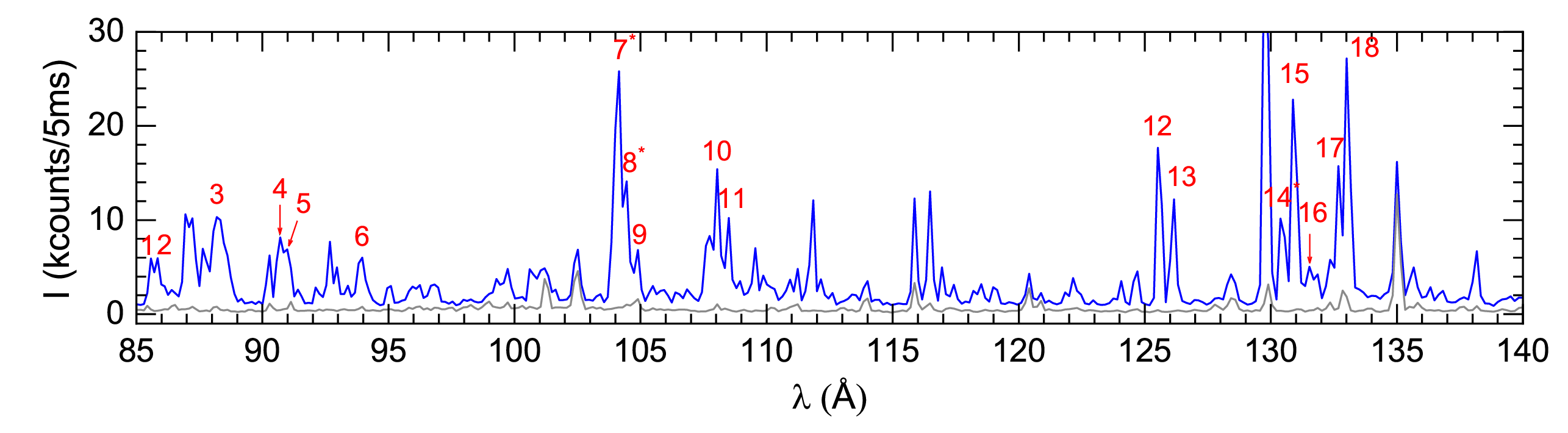}   
\caption{EAST spectrum and the lines identified as belonging to Al$^{4+}$. 
Gray curve: spectrum before the Al burst. Blue curve: spectra at maximum Al emission. 
The numbers indicate the lines (Key) described in Table~\ref{table:Al4}.}
\label{fig:Al4}
\end{figure}                      
\end{center}

\begin{table}[h!]
\caption{Description of the theoretical model used for the calculation of 
Al$^{4+}$ spectra. $I_{\mathrm{p}}$: Ionization Energy. 
$T_{\mathrm{max}}$: Temperature of maximum abundance. 
$N$: Number of levels considered in CRM, 
$N_{E1}$: Number of dipole-allowed radiative transitions. }
\begin{indented}
\item[]\begin{tabular}{ccccc}
\br
\multicolumn{1}{c|}{Ground Configuration} & 
\multicolumn{1}{c|}{$I_{\mathrm{p}}$ (eV)} & 
\multicolumn{1}{c|}{$T_{\mathrm{max}}$ (eV)} & 
\multicolumn{1}{c|}{$N$}  &
\multicolumn{1}{c}{$N_{E1}$}    \\
\cline{1-5}
\multicolumn{1}{c|}{$1s^{2}2s^{2}2p^{5}$ (F-like)}      &
\multicolumn{1}{c|}{152}      & 
\multicolumn{1}{c|}{20}      & 
\multicolumn{1}{c|}{740}   &
\multicolumn{1}{c}{74247}   \\ 
\cline{1-5}  
\mr
\multicolumn{1}{l}{Odd Configurations:} &
\multicolumn{4}{l}{  
$1s^{2}2s^{2}2p^{5}$,   
$1s^{2}2s^{2}2p^{4}nl$ $(n=3,4,5 ~;~ l=1,3)$, }  \\
\multicolumn{1}{l}{} &
\multicolumn{4}{l}{ 
$1s^{2}2s^{2}2p^{3}3l^{2}$ $(l=0,1,2)$, 
$1s^{2}2s2p^{5}nl$  ($n=3,4,5 ~;~ l=0,2,4)$ }   \\
\cline{1-5}
\multicolumn{1}{l}{Even Configurations:} &
\multicolumn{4}{l}{  
$1s^{2}2s^{2}2p^{4}nl$ $(n=3,4,5 ~;~ l=0,2,4)$, }  \\
\multicolumn{1}{l}{} &
\multicolumn{4}{l}{ 
$1s^{2}2s^{2}2p^{3}3s3p$, $2s2p^{6}$, 
$1s^{2}2s2p^{5}nl$ $(n=3,4,5 ~;~ l=1,3)$ }   \\
\br
\label{table:Al4res}
\end{tabular}
\end{indented}
\end{table}
As in the previous case, the transitions with the highest radiative rate coefficients do not necessarily dominate the spectrum. 
Here, the transitions $2s^22p^33d^2$ to $2s^22p^43d$ (i.e., $3d \rightarrow 2p)$ and $2s2p^54d$ to $2s2p^6$ $(4d \rightarrow 2p)$ have the largest Einstein coefficients, of the order of $10^{11}$ s$^{-1}$. Lines belonging to the first group should appear at 90.1 \AA~and 94.2 \AA, and to the second, around 96.8 \AA. However, our collisional-radiative model shows that the upper levels of these transitions are not populated. 
Indeed, as is shown in Fig.~\ref{fig:Al4}, these lines are hardly seen in the spectrum. 
This Figure shows the region of the EAST spectra in which it is possible to identify 18 lines corresponding to Al$^{4+}$ ion transitions.
These lines are described in Table~\ref{table:Al4}, and again, the experimental results are compared with EBIT results, NIST experimental tabulated data, and our theoretical calculation results given by  {\sc fac}, {\sc hullac}, and {\sc as}.

Line identification is more difficult than other cases, because different transitions overlap in the same region.  
Table~\ref{table:Al4} shows that most of the prominent lines correspond to decay to the ground configuration.
The Al$^{4+}$ spectrum consists of four easily identifiable regions. 
The first one (between 85 and 95 \AA) contains the $nd \rightarrow 2p$ transitions listed in the Table as 1-6 lines. 
These lines are blended with a similar group of lines belonging to the Al$^{5+}$ ion.
A noticeable structure of three pronounced peaks between 104 and 105 \AA~corresponds to the 
$4d \rightarrow 2p$ transitions (lines 7$^*$, 8$^*$, and 9 in Table~\ref{table:Al4}). Although our calculations allow line identification, we obtained the same features for the Al$^{5+}$ ion, overlapping at the same place (lines 5$^*$ and 6$^*$ in Table~\ref{table:Al5}). 
The third group of lines has the $3d \rightarrow 2p$ transitions (10-11), which are strong but not very well isolated.
The last group of Al$^{4+}$ lines covers the region 125-133 \AA~and is formed by $3s \rightarrow 2p$ lines (lines 12 to 18).
These are striking and identifiable lines.
As in the previous case, we found good agreement between the theoretical predictions and the experimental results. Again, {\sc fac} wavelengths are slightly higher than the experimental values, whereas {\sc as} results are below the measurements. Also, the {\sc hullac} results are in excellent agreement with the experimental values, about in the 0.5\% range of discrepancy.
The agreement between our results and EBIT values is excellent, except for lines 1, 3, and 17, where the differences are slightly higher than $\Delta \lambda=0.2$ \AA. However, for the last line, it is worth mentioning that NIST reported the wavelength value $\lambda=132.626$ \AA~from \cite{Artru:74},  which differs by only 0.05 \AA~from our experimental value.

\begin{table}[h!]
\caption{EAST experimental and theoretical lines ({\sc fac}, {\sc hullac}, and {\sc autostructure}), for Al$^{4+}$.
The values from other experiments and the experimental wavelengths quoted in 
NIST are also displayed.
The column `Key' indicates the transitions labeled in Fig.~\ref{fig:Al4}.
The symbol $^*$ means that the line is blended with another ion's line, 
as explained in the text. 
We use the usual relativistic nomenclature $j_{\pm} \equiv l \pm \frac{1}{2}$. 
Wavelengths are in \AA.}
\label{table:Al4}
\footnotesize
\begin{tabular}{lcccccccc}
\br
Key & Exp. & {\sc fac} & {\sc hullac} & AS & Previous Exp. & NIST & Lower  &  Upper \\
\mr
1 & 85.58 & 86.28 & 85.59 & 85.30 & $85.817^a$ & 85.804 & $2p^5_{+}$ & $(2p_{-}2p^3_{+}5d_{-})_{3/2}$  \\
2 & 85.85 & 86.50 & 85.84 & 85.39 &  & 85.922 & $2p^5_{-}$ & $(2p_{-}2p^3_{+})5d_{+})_{3/2}$  \\       
3 & 88.20 & 89.27 & 88.67 & 87.46 & $88.480^a$ & 88.539 & $2p^5_{-}$ & $(2p^4_{+}5d_{-})_{3/2}$ \\           
4 & 90.71 & 91.34 & 90.58 & 90.22 & $90.889^a$ & 90.701 & $2p^5_{+}$ & $(2p_{-}2p^3_{+}4d_{-})_{5/2}$ \\ 
5 & 90.99 & 91.42 & 90.85 & 90.47 &  & 90.982 & $2p^5_{+}$ & $(2p_{-}2p^3_{+}4d_{+})_{3/2}$ \\                 
6 & 93.96 & 94.41 & 93.75 & 94.59 & $94.024^a$ & 93.755 & $2p^5_{+}$ & $(2p_{-}2p^3_{+}4d_{-})_{5/2}$ \\  
7$^*$ & 104.15 & 104.84 & 104.01 & 103.27 & $104.196^a$ & 104.072 & $2p^5_{-}$ & $(2p_{-}2p^3_{+}3d_{+})_{1/2}$ \\ 
8$^*$& 104.45 & 105.13 & 104.10 & 103.57 &  & 104.362 & $2p^5_{-}$ & $(2p_{-}2p^3_{+}3d_{-})_{1/2}$  \\ 
9  & 104.89 & 105.13 & 104.12 & 103.86 &  &   & $2p^5_{-}$ & $(2p_{-}2p^3_{+}3d_{+})_{3/2}$ \\
10 & 108.05 & 109.14 & 108.17 & 108.18  & $108.072^a$ & 108.005 & $2p^5_{+}$ & $(2p_{-}2p^3_{+}3d_{+})_{5/2}$ \\	 
11 & 108.50 & 109.15 & 108.91 & 108.39  & $108.587^a$ & 108.462 & $2p^5_{-}$ & $(2p^2_{-}2p^2_{+}3d_{-})_{3/2}$ \\ 
12 & 125.52 & 126.80 & 125.14 & 126.65 & $125.598^a$ & 125.529 & $2p^5_{+}$ & $(2p_{-}2p^3_{+}3s)_{5/2}$ \\ 
13 & 126.16 & 127.33 & 125.65 & 127.10& $126.140^a$ $126.05^b$& 126.068 & $2p^5_{-}$ & $(2p_{-}2p^3_{+}3s)_{3/2}$ \\  
$14^*$ & 130.39 & 131.33 &   &    & $130.475^a$ & 130.390 & $2p^5_{+}$ & 
$(2p^2_{-}2p^2_{+}3s)_{3/2}$  \\
15 & 130.88 & 131.79 & 130.75 & 129.14 & $130.944^a$ $130.848^b$& 131.002 & $2p^5_{+}$ & $(2p_{-}2p^3_{+}3s)_{1/2}$ \\   
16 & 131.53 & 131.92 & 131.17 & 129.52 & $131.532^a$ & 131.438 & $2p^5_{-}$ & $(2p^2_{-}2p^2_{+}3s)_{3/2}$ \\
17 & 132.68 & 132.38 & 132.74 & 130.15 & $132.964^a$ & 132.626 & $2p^5_{+}$ & $(2p_{-}2p^3_{+}3s)_{3/2}$ \\ 
18 & 133.01 & 134.24 & 133.09 & 130.56 &   & 133.233 & $2p^5_{+}$ & $(2p^4_{+}3s)_{1/2}$    \\
\br    
\end{tabular}
\begin{indented}
\item[] $^a$ Livermore EBIT experimental results \cite{Gu11}.
\item[] $^b$ Laser-produced plasma experimental results \cite{Valero72}.
\end{indented}
\end{table}

\clearpage
\subsection{Al$^{5+}$}

The Al$^{5+}$ ion is isoelectronic to O; its ground configuration is 
$1s^22s^22p^4$. The first ionization limit is 188.8 eV, and the maximum fractional abundance of this ion is around 30 eV.
For the theoretical calculations of Al$^{5+}$, we include in our model 
the thousand of levels belonging to the configurations listed in 
Table~\ref{table:Al5res}.
All the configuration interaction mixing among them are taking into 
account. 
Only the electron-dipole allowed ones have been considered from a total of $6 \times 10^5$ transitions, resulting in more than 150000 radiative rate coefficients. 
\begin{table}[h!]
\caption{Description of the theoretical model used for the calculation of 
Al$^{5+}$ spectra. $I_{\mathrm{p}}$: Ionization Energy. 
$T_{\mathrm{max}}$: Temperature of maximum abundance. 
$N$: Number of levels considered in CRM, 
$N_{E1}$: Number of dipole-allowed radiative transitions. }
\begin{indented}
\item[]\begin{tabular}{ccccc}
\br
\multicolumn{1}{c|}{Ground Configuration} & 
\multicolumn{1}{c|}{$I_{\mathrm{p}}$ (eV)} & 
\multicolumn{1}{c|}{$T_{\mathrm{max}}$ (eV)} & 
\multicolumn{1}{c|}{$N$}  &
\multicolumn{1}{c}{$N_{E1}$}    \\
\cline{1-5}
\multicolumn{1}{c|}{$1s^{2}2s^{2}2p^{4}$ (O-like)}      &
\multicolumn{1}{c|}{189}      & 
\multicolumn{1}{c|}{30}      & 
\multicolumn{1}{c|}{1088}   &
\multicolumn{1}{c}{152729}   \\ 
\cline{1-5}  
\mr
\multicolumn{1}{l}{Even Configurations:} &
\multicolumn{4}{l}{  
$1s^{2}2s^{2}2p^{4}$,   
$1s^{2}2s^{2}2p^{3}nl$ $(n=3,4,5 ~;~ l=1,3)$, }  \\
\multicolumn{1}{l}{} &
\multicolumn{4}{l}{ 
$1s^{2}2s^{2}2p^{2}3l^{2}$ $(l=0,1,2)$, 
$1s^{2}2s2p^{4}nl$  ($n=3,4,5 ~;~ l=0,2,4)$ }   \\
\cline{1-5}
\multicolumn{1}{l}{Odd Configurations:} &
\multicolumn{4}{l}{  
$1s^{2}2s^{2}2p^{3}nl$ $(n=3,4,5 ~;~ l=0,2,4)$, }  \\
\multicolumn{1}{l}{} &
\multicolumn{4}{l}{ 
$1s^{2}2s^{2}2p^{2}3s3p$, $2s2p^{5}$, 
$1s^{2}2s2p^{4}nl$ $(n=3,4,5 ~;~ l=1,3)$ }   \\
\br
\label{table:Al5res}
\end{tabular}
\end{indented}
\end{table}

\begin{center}
\begin{figure}[h!]
\centering\includegraphics[width=0.95\textwidth]{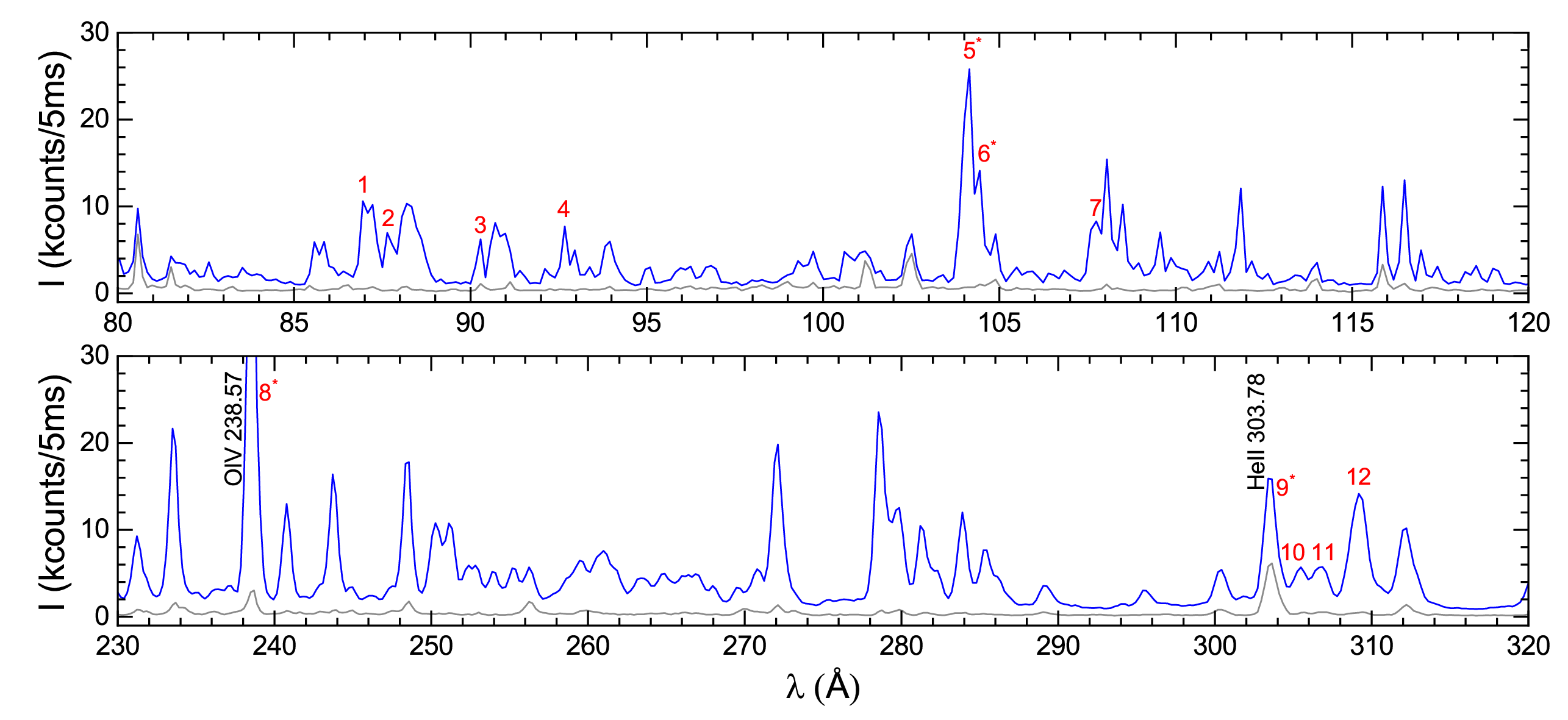}   
\caption{EAST spectrum and the lines identified as belonging to Al$^{5+}$.
Gray curve: spectrum before the Al burst. Blue curve: spectra at maximum Al emission. 
The numbers indicate the lines (Key) described in Table~\ref{table:Al5}. }
\label{fig:Al5}
\end{figure}                      
\end{center}

In this system, the transitions with the highest radiative rate coefficients are the 
$2s^{2}2p^{2}3d^{2} \rightarrow 2s^{2}2p^{3}3d$,  with wavelengths close to the 75-78.5 \AA, and Einstein coefficient values of about $5\times10^{11}$ s$^{-1}$.
However, the lines appearing in that region of the experimental spectra belong to the Al$^{6+}$ ion. 
The EAST spectrum is shown in  Fig.~\ref{fig:Al5}, and the experimental and theoretical results for 12 lines of this ion are shown in Table~\ref{table:Al5}. 
The spectrum of Al$^{5+}$ ion consists of a first group of lines located between 87 \AA~and 93 \AA. 
Many of these lines belong to the $2s^{2}2p^{3}3d  \rightarrow 2s^{2}2p^{4}$ 
transitions, but only some of them can be identified with certainty (lines 1 to 4 in Table~\ref{table:Al5}).
The next group of lines, between  102 \AA~and 110 \AA, corresponds to the 
$2s^{2}2p^{3}3s  \rightarrow 2s^{2}2p^{4}$ transitions. 
As pointed out above, the Al$^{4+}$ and Al$^{5+}$ lines coexist at this region, and the calculations of both ions reproduced the same features. 
Next, we found theoretically a huge line at 236 \AA, corresponding to the transition from the 
$2s^2p^{4}$ to the fourth level of the ground configuration ($2p \rightarrow 2s$ transition), which is missing in the experimental spectrum.
We included more configuration interactions in the calculations, adding $2s^22p^{3}nl$ and $2s2p^{4}nl$ $(n=6,7)$ to investigate the possibility of shifting the position of the lines towards one of the closest experimental peaks.  
However, the changes are not relevant for this line identification. 
We do not find any explanation that can account for the disappearance of this line. Therefore, we identify it as the large line seen in the spectrum at 238.48 \AA~(line 8$^*$), which also corresponds to an O IV line at 238.57 \AA. 
Our CR model identifies as the most important transitions, the $2s^{2}2p^{4} \rightarrow 2s2p^{5}$, with wavelengths in the range 300-310 \AA. 
The convolution of the many lines appearing at this region results in five peaks, and the agreement with the experimental values is very good, better than 0.3 \%. 
However, our CR model does not account very well for the relative intensities of the peaks. 
The strongest theoretical ({\sc hullac} line results in 306.4 \AA~(line 10), which is not the higher peak of this group. 
Our calculations find the line 9$^*$ ($2p \rightarrow 2s$ transition) at 304.2 \AA, the same position as a $2s2p^5 \rightarrow 2s^22p^4$ transition of Al$^{6+}$ (line 14$^*$ of Table~\ref{table:Al6}). 
As displayed with a fine gray line, they overlap with a line of HeII (at 303.78 \AA), which can explain the high-intensity peak in the experimental spectra.

The {\sc as} results for the spectrum at this region have been obtained in a separate calculation without mixing with higher-$n$ configurations. The agreement with {\sc hullac} results is remarkable.
We included in Table~\ref{table:Al5} the measured values of 305.05 \AA~and 309.09 \AA~observed in the TJ-II stellarator plasma experiments of 
McCarthy {\it et al.} \cite{McCarthy16}. 
However, they assigned these lines to the Al$^{8+}$ and Al$^{6+}$ ions, respectively. 
Concerning the first line, we found no relevant Al$^{8+}$ 
line in the range 286-312 \AA. 
Regarding the second line, our three independent calculations 
encounter the closest Al$^{6+}$ line at 303 \AA, which seems too 
far out to the 309 \AA~wavelength.
The other experimental value marked in red is the 239.030 \AA~line obtained in a laser-produced plasma experiment by Valero and Goorvitch  \cite{Valero72}, assigned by these authors to Al$^{6+}$.
In our calculations, we have not found any transition that could contribute significantly to the Al$^{6+}$ spectrum in the region 233-259 \AA. 
Therefore, although we have a 1\% difference from the experimental results, we rule out that this line comes from the Al$^{6+}$ emission.
The agreement with EBIT results is excellent. 
We have a large discrepancy (0.38 \AA) with NIST results in line 1 and even more significant for the lines beyond 300 \AA~(11 and 12). 

\begin{table}[h!]
\caption{EAST experimental and theoretical lines for Al$^{5+}$.
The column `Key' indicates the transitions labeled in Fig.~\ref{fig:Al5}.
The reference values in red are lines assigned by the authors to other Al ions.
We use the usual relativistic nomenclature $j_{\pm} \equiv l \pm \frac{1}{2}$. 
The values shown in red refer to lines assigned to other ions.     
Wavelengths are in \AA.}
\label{table:Al5}
\footnotesize
\begin{tabular}{lcccccccc}
\br
Key & Exp. & {\sc fac} & {\sc hullac} & AS & Previous Exp. & NIST & Lower & Upper \\
\mr
1  & 86.95 & 87.67 & 87.07 & 84.60 &   & 87.331 & $(2p_{-}2p^3_{+})_2$ & $\left(2p^3_{+}3d_{-}\right)_3$  \\        
2  & 87.64 & 87.88 & 87.71 & 85.31 & $87.760^a$ & 87.651 & $(2p^2_{-}2p^2_{+})_2$ & $\left(2p_{-}2p^2_{+}3d_{-} \right)_2$  \\ 
3  & 90.29 & 90.54 & 89.73 & 86.82 & $90.310^a$ $90.200^b$ & 90.196 & $(2p_{-}2p^3_{+})_2$ & $\left(2p^2_{-}2p_{+}3d_{-}\right)_3$  \\  
4  & 92.68 & 92.88 & 92.27 & 89.00 & $92.833^a$ & 92.626 & $(2p^2_{-}2p^2_{+})_2$ & $\left(2p_{-}2p^2_{+}3d_{+}\right)_2$  \\      
5$^*$ & 104.15 & 104.55 & 103.58 & 100.71 & $104.196^a$ & 104.046 & $(2p^2_{-}2p^2_{+})_2$ & $\left(2p_{-}2p^2_{+}3s\right)_3$  \\ 
6$^*$ & 104.45 & 104.84 & 103.86 & 101.04 &  & 104.464 & $(2p_{-}2p^3_{+})_1$ & $\left(2p^2_{-}2p_{+}3s \right)_2$  \\ 
7  & 107.74 & 108.27 & 107.29 & 104.08 &    & 107.622 & $(2p_{-}2p^3_{+})_2$ & $\left(2p_{-}2p^2_{+}3s\right)_2$  \\
8$^*$ & 238.48 & 235.83 & 236.07 & 237.09 & \color {red} $239.030^b$ &   & $(2s^22p_{-}2p^3_{+})_2$  & $\left(2s2p^2_{-}2p^3_{+}\right)_1$  \\
9$^*$ & 303.55 & 303.34 & 304.20 & 303.60 &    &   & $(2s^22p^2_{-}2p^2_{+})_2$  & $\left(2s2p_{-}2p^4_{+}\right)_1$  \\ 
10 & 305.49 & 304.56 & 306.40 & 306.29 & \color {red} $305.05^c$&   & $(2s^22p^2_{-}2p^2_{+})_2$  & $\left(2s2p^2_{-}2p^3_{+}\right)_2$  \\ 
11  & 306.64 & 305.59 & 307.63 & 307.63 & $307.248^b$ & 307.249 & $(2s^22p^4_{+})_0$  & $\left(2s2p_{-}2p^4_{+}\right)_1$ \\ 
12  & 309.19 & 306.85 & 308.74 & 309.16 & \color {red} $309.09^c$ & 309.597 & $(2s^22p_{-}2p^3_{+})_1$  & $\left(2s2p^2_{-}2p^3_{+}\right)_2$  \\
\br  
\end{tabular}
\begin{indented}
\item[] $^a$ Livermore EBIT experimental results \cite{Gu11}.
\item[] $^b$ Laser-produced plasma experimental results \cite{Valero72}.
\item[] $^c$ TJ-II stellarator plasma experimental results \cite{McCarthy16}.
\end{indented}
\end{table}

\clearpage
\subsection{Al$^{6+}$}

This ion, isoelectronic to the N, has five levels in the ground 
configuration $1s^22s^22p^{3}$, separated by 14 eV.
The configurations included in the calculation are listed in 
Table~\ref{table:Al6res}. 
All mixing between these configurations have been taken into account.
\begin{table}[h!]
\caption{Description of the theoretical model used for the calculation of 
Al$^{6+}$ spectra. $I_{\mathrm{p}}$: Ionization Energy. 
$T_{\mathrm{max}}$: Temperature of maximum abundance. 
$N$: Number of levels considered in CRM, 
$N_{E1}$: Number of dipole-allowed radiative transitions. }
\begin{indented}
\item[]\begin{tabular}{ccccc}
\br
\multicolumn{1}{c|}{Ground Configuration} & 
\multicolumn{1}{c|}{$I_{\mathrm{p}}$ (eV)} & 
\multicolumn{1}{c|}{$T_{\mathrm{max}}$ (eV)} & 
\multicolumn{1}{c|}{$N$}  &
\multicolumn{1}{c}{$N_{E1}$}    \\
\cline{1-5}
\multicolumn{1}{c|}{$1s^{2}2s^{2}2p^{3}$ (N-like)}      &
\multicolumn{1}{c|}{242}      & 
\multicolumn{1}{c|}{45}      & 
\multicolumn{1}{c|}{982}   &
\multicolumn{1}{c}{132273}   \\ 
\cline{1-5}  
\mr
\multicolumn{1}{l}{Odd Configurations:} &
\multicolumn{4}{l}{  
$1s^{2}2s^{2}2p^{3}$,   
$1s^{2}2s^{2}2p^{2}nl$ $(n=3,4,5 ~;~ l=1,3)$, }  \\
\multicolumn{1}{l}{} &
\multicolumn{4}{l}{ 
$1s^{2}2s^{2}2p3l^{2}$ $(l=0,1)$, 
$1s^{2}2s2p^{3}nl$  ($n=3,4,5 ~;~ l=0,2,4)$ }   \\
\cline{1-5}
\multicolumn{1}{l}{Even Configurations:} &
\multicolumn{4}{l}{  
$1s^{2}2s^{2}2p^{2}nl$ $(n=3,4,5 ~;~ l=0,2,4)$, }  \\
\multicolumn{1}{l}{} &
\multicolumn{4}{l}{ 
$1s^{2}2s^{2}2p3s3p$, $2s2p^{4}$, 
$1s^{2}2s2p^{3}nl$ $(n=3,4,5 ~;~ l=1,3)$ }   \\
\br
\label{table:Al6res}
\end{tabular}
\end{indented}
\end{table}

\begin{center}
\begin{figure}[h!]
\centering\includegraphics[width=0.95\textwidth]{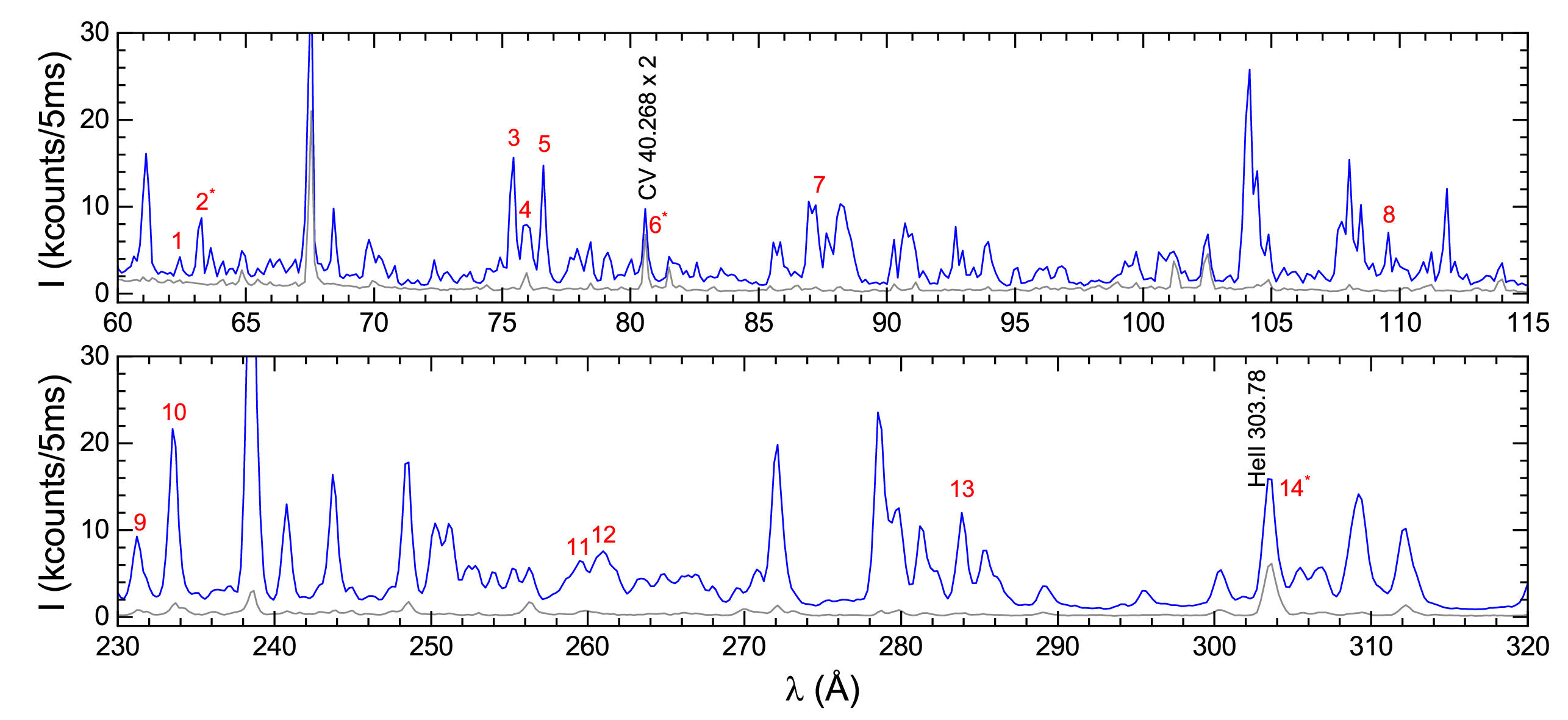}   
\caption{EAST spectrum and the lines identified as belonging to Al$^{6+}$.
Gray curve: spectrum before the Al burst. Blue curve: spectra at maximum Al emission. 
The numbers indicate the lines (Key) described in Table~\ref{table:Al6}.}
\label{fig:Al6}
\end{figure}                      
\end{center}

More than 500 transitions have radiative transition rates higher than $10^{11}$ s$^{-1}$ in the range of wavelengths observed at EAST.
The higher of them are 
$2s^{2}2p^{2}3d \rightarrow 2s^22p^3$,  with wavelengths close to 74  \AA~-- 76 \AA, 
and Einstein coefficient values of about $5\times10^{11}$ s$^{-1}$.
The Al$^{6+}$ spectrum, shown in Fig.~\ref{fig:Al6}, consists of a broad region between 60 \AA~and 90 \AA, with many lines corresponding to transitions from $2p^23s$, $2p^23d$, $2p^24s$, and $2p^24d$ to the ground $2p^3$ levels. 
There is another broad region, between 230 \AA~and 305 \AA, in which we can identify strong lines 
belonging to the $2s2p^4 \rightarrow 2s^22p^3$ transitions. 
The most prominent line arising in the theoretical model is the line labeled in Table~\ref{table:Al6} as line 14$^*$, whose calculated wavelength is 303.5 \AA~({\sc hullac}). 
As commented above, the 303.78 \AA~line identified as belonging to the HeII ion also appears at this wavelength. 
The other blended lines are the 2$^*$ at 63.27 \AA, which overlaps with the 3$^*$ line of Al$^{8+}$ (Table~\ref{table:Al8}), and the 6$^*$ at 80.04 \AA, which overlaps with the 4$^*$ line of Al$^{7+}$ (Table~\ref{table:Al7}).
Line 13 observed at 283.88 \AA~is predicted by {\sc fac} and {\sc as} 
calculations within the 0.1 \% of error. On the other hand, our {\sc hullac} 
calculation produces a wavelength value that is  3 \AA~higher. 
In the three cases, the CR models predict a strong intensity line.
However, this line is melded with one belonging to the Al$^{8+}$ ion,  
as shown in Table \ref{table:Al8}. The laser-produced plasma experiments by Valero and Goorvitch \cite{Valero72} also found an Al$^{8+}$ line at 282.634 \AA, and McCarthy {\it et al.} \cite{McCarthy16} reported an Al$^{8+}$ line at 284.04 \AA.
The value reported at NIST is from Valero  \cite{Valero:75}, but the 
transition is adjudicated differently.

\begin{table}[h!]
\caption{EAST experimental and theoretical lines for Al$^{6+}$.
The column `Key' indicates the transitions labeled in Fig.~\ref{fig:Al6}.
We use the usual relativistic nomenclature $j_{\pm} \equiv l \pm \frac{1}{2}$. 
The values shown in red refer to lines assigned to other ions.     
Wavelengths are in \AA.}
\label{table:Al6}
\footnotesize
\begin{tabular}{lcccccccc}
\br
Key & Exp. & {\sc fac} & {\sc hullac} & AS & Previous Exp. & NIST & Lower & Upper \\
\mr
1 & 62.42  & 62.03 & 61.82 & 60.25 & $62.380^a$ & 62.292 & $(2p_{-}2p^2_{+})_{3/2}$ & $(2p^2_{-}4d_{+})_{5/2}$ \\
2$^*$ & 63.27  & 64.52 & 62.69 & 61.16 & $63.161^a$ & 63.056 & $(2p_{-}2p^2_{+})_{5/2}$ & $(2p_{-}2p_{+}4d_{+})_{7/2}$ \\
3 & 75.43  & 74.89 & 74.55 & 72.39 & $75.434^a$ & 75.367 & $(2p_{-}2p^2_{+})_{3/2}$ & $(2p^2_{-}3d_{+})_{5/2}$ \\
4 & 75.95  & 75.13 & 75.16 & 72.71 &   & 75.903 & $(2p_{-}2p^2_{+})_{1/2}$ & $(2p_{-}2p_{+}3d_{+})_{3/2}$ \\
5 & 76.61  & 76.18 & 75.80 & 73.51 & $76.644^a$ & 76.582 & $(2p_{-}2p^2_{+})_{5/2}$ & $\left(2p_{-}2p_{+}3d_{+}\right)_{7/2}$ \\
6$^*$ & 80.04  & 80.35 & 80.03 & 77.17 & $80.043^a$ & 79.928 & $(2p^3_{+})_{3/2}$ & $\left(2p_{-}2p_{+}3d_{+}\right)_{5/2}$ \\
7 & 87.23 & 87.07 & 87.25 & 86.64 & $87.092^a$ & 87.165 &$(2p_{-}2p^2_{+})_{3/2}$ & $\left(2p_{-}2p_{+}3s\right)_{3/2}$ \\
8 & 109.56 & 110.29 & 109.73 & 107.60 &  &  &$\left(2s^22p_{-}2p_{+}3p_{+}\right)_{5/2}$ & $\left(2s2p_{-}2p^3_{+}\right)_{5/2}$ \\
9 & 231.23 & 229.84 & 230.73 & 231.06 &  &  & $(2s^22p^2_{-}2p_{+})_{3/2}$ & $\left(2s2p_{-}2p^3_{+}\right)_{1/2}$ \\
10 & 233.50 & 231.37 & 232.27 & 232.95 &  &  & $(2s^22p_{-}2p^2_{+})_{5/2}$ & $(2s2p^2_{-}2p^2_{+})_{3/2}$ \\
11 & 259.48 & 257.16 & 258.81 & 258.58 & $259.035^b$ & 259.207 & $(2s^22p^3_{+})_{3/2}$ & $\left(2s2p_{-}2p^3_{+}\right)_{1/2}$ \\
12 & 260.99 & 259.24 & 260.66 & 260.77 &  & 261.044 & $(2s^22p^3_{+})_{3/2}$ & $(2s2p^2_{-}2p^2_{+})_{3/2}$ \\
13$^*$ & 283.88 & 284.09 & 286.86 & 284.24 & \color{red} $282.634^b$ \color{red} $284.04^c$& 282.66 & $(2s^22p^3_{+})_{3/2}$ & $(2s2p^2_{-}2p^2_{+})_{1/2}$ \\
14$^*$ & 303.55 & 303.52 & 303.48 & 300.27 &  &  & $(2s^22p_{-}2p^2_{+})_{5/2}$ & $\left(2s2p_{-}2p^3_{+}\right)_{5/2}$ \\
\br
\end{tabular}
\begin{indented}
\item[] $^a$ Livermore EBIT experimental results \cite{Gu11}.
\item[] $^b$ Laser-produced plasma experimental results \cite{Valero72}.
\item[] $^c$ TJ-II stellarator plasma experimental results \cite{McCarthy16}.
\end{indented}
\end{table}

\clearpage
\subsection{Al$^{7+}$}
\label{subsec:Al7}

This ion is isoelectronic to the C, having a ground configuration $1s^22s^22p^2$. 
The ionization energy is 284.6 eV, and the maximum fractional abundance is about 60 eV. 
The configurations included in the calculation are listed in 
Table~\ref{table:Al7res}. 
As in all the other cases, all the configuration mixing are 
considered.
\begin{table}[h!]
\caption{Description of the theoretical model used for the calculation of 
Al$^{7+}$ spectra. $I_{\mathrm{p}}$: Ionization Energy. 
$T_{\mathrm{max}}$: Temperature of maximum abundance. 
$N$: Number of levels considered in CRM, 
$N_{E1}$: Number of dipole-allowed radiative transitions. }
\begin{indented}
\item[]\begin{tabular}{ccccc}
\br
\multicolumn{1}{c|}{Ground Configuration} & 
\multicolumn{1}{c|}{$I_{\mathrm{p}}$ (eV)} & 
\multicolumn{1}{c|}{$T_{\mathrm{max}}$ (eV)} & 
\multicolumn{1}{c|}{$N$}  &
\multicolumn{1}{c}{$N_{E1}$}    \\
\cline{1-5}
\multicolumn{1}{c|}{$1s^{2}2s^{2}2p^{2}$ (C-like)}      &
\multicolumn{1}{c|}{285}      & 
\multicolumn{1}{c|}{45}      & 
\multicolumn{1}{c|}{833}   &
\multicolumn{1}{c}{88270}   \\ 
\cline{1-5}  
\mr
\multicolumn{1}{l}{Even Configurations:} &
\multicolumn{4}{l}{  
$1s^{2}2s^{2}2p^{2}$,   
$1s^{2}2s^{2}2pnl$ $(n=3,4,5 ~;~ l=1,3)$, }  \\
\multicolumn{1}{l}{} &
\multicolumn{4}{l}{ 
$1s^{2}2s2p^{2}nl$  ($n=3,4,5 ~;~ l=0,2,4)$ }   \\
\cline{1-5}
\multicolumn{1}{l}{Odd Configurations:} &
\multicolumn{4}{l}{  
$1s^{2}2s^{2}2pnl$ $(n=3,4,5 ~;~ l=0,2,4)$, 
$1s^{2}2s^{2}2p3l^{2}$ $(l=0,1,2)$, }  \\
\multicolumn{1}{l}{} &
\multicolumn{4}{l}{ 
$1s^{2}2s2p^{3}$, 
$1s^{2}2s2p^{2}nl$ $(n=3,4,5 ~;~ l=1,3)$ }   \\
\br
\label{table:Al7res}
\end{tabular}
\end{indented}
\end{table}

\begin{center}
\begin{figure}[h!]
\centering\includegraphics[width=0.95\textwidth]{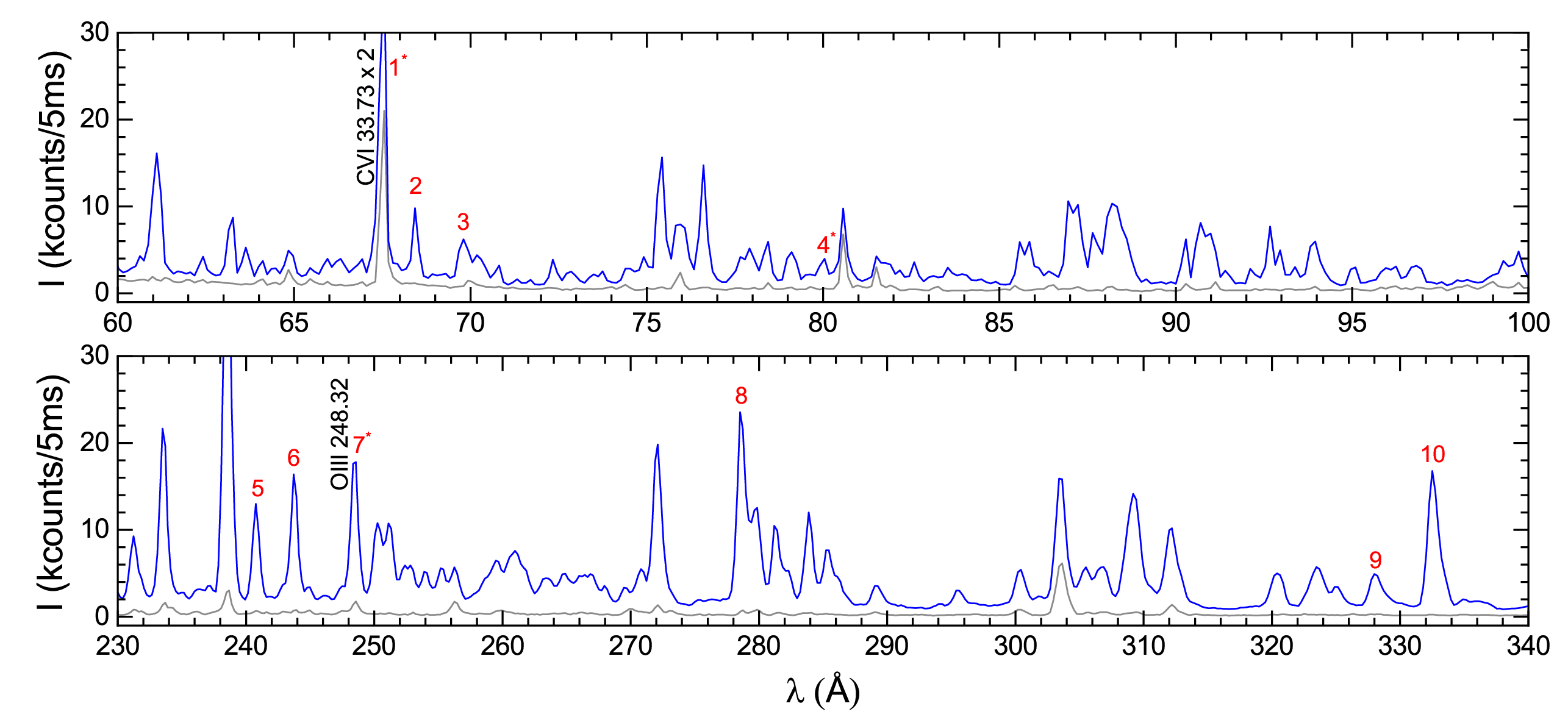}   
\caption{EAST spectrum and the lines identified as belonging to Al$^{7+}$.
Gray curve: spectrum before the Al burst. Blue curve: spectra at maximum Al emission. 
The numbers indicate the lines (Key) described in Table~\ref{table:Al7}.}
\label{fig:Al7}
\end{figure}                      
\end{center}

Identifying the Al$^{7+}$ spectrum is complicated due to the large number of transitions (more than 180) with values above $10^{12}$ s$^{-1}$.
However, as is shown in Fig.~\ref{fig:Al7}, only a few strong lines can be assigned to this ion.
The CR model allows us to identify the dominant lines listed in the Table~\ref{table:Al7}. 
Most of the strong transitions are in the range 50 \AA~to 75 \AA, mainly corresponding to the 
$2s2p3d^2 \rightarrow 2s2p^{2}3d$ and $2p3d \rightarrow 2p^{2}$ transitions. 
There are also prominent lines corresponding to the $2s2p^{3} \rightarrow 2s^22p^2$,  
with wavelengths in the range 235-330 \AA, particularly lines 8 and 10. 
We have added $n=6$ and $n=7$ to investigate the effect of mixing configurations in improving the theoretical model, but this has not produced significant changes in the results. 
For example, the theoretical value obtained for the strong line 10 by a {\sc hullac} calculation including only $n=3,4,5$ is 335.8 \AA, while the improvement obtained by adding $n=6,7$ (335.5 \AA) is still insufficient to bring the value closer to the experimental result of 332.51 \AA. 
It is noticeable that this line has also been identified in a stellarator plasma by McCarthy {\it et al.}  \cite{McCarthy16} and also in NIST, but as belonging to the Al$^{9+}$ ion.
We included in Table~\ref{table:Al7} other lines observed in laser-produced plasma experiments by Valero and Goorvitch  \cite{Valero72}, although most of them are labeled in red since the authors assign these lines to other ions. According to them, line 240.770 \AA~corresponds to Al$^{6+}$, line 243.760 \AA~corresponds to Al$^{5+}$, and line 278.699 \AA~corresponds to the Al$^{4+}$ ion. Our calculations can not reproduce any value close to these lines for these 
ions.  
We have marked with an asterisk, in addition to line 4$^*$, lines 1$^*$ and 7$^*$ because in the first case, the peak overlaps with a high-intensity second-order C VI line ($2 \times 33.73$  \AA), and in the second, the 7$^*$ line overlaps with an O III line at 248.32 \AA.

\begin{table}[h!]
\caption{EAST experimental and theoretical lines for Al$^{7+}$.
The column `Key' indicates the transitions labeled in Fig.~\ref{fig:Al7}.
We use the usual relativistic nomenclature $j_{\pm} \equiv l \pm \frac{1}{2}$. 
The values shown in red refer to lines assigned to other ions.     
Wavelengths are in \AA.}
\label{table:Al7}
\footnotesize
\begin{tabular}{lcccccccc}
\br
Key & Exp. & {\sc fac} & {\sc hullac} & AS & Previous Exp. & NIST & Lower & Upper \\
\mr                                         
1$^*$ & 67.68 & 67.51 & 66.92 & 65.43 & $67.440^a$ & 67.946 & $(2p^2_{+})_2$ & $(2p_{+}3d_{+})_3$ \\  
2 & 68.43 & 68.41 & 67.81 & 66.15 & $68.446^a$ $68.375^b$& 68.375 & $(2p_{-}2p_{+})_2$ & $(2p_{+}3d_{-})_3$ \\  
3 & 69.81 & 70.39 & 69.75 & 69.45 & $69.726^a$ & 70.161 & $(2p_{-}2p_{+})_2$ & $(2p_{+}3d_{-})_2$ \\               
4$^*$ & 80.04 & 80.55 & 80.03 & 80.11 & $80.514^a$ & 80.320 & $(2s2p_{-}2p^2_{+})_3$ & $(2s2p^2_{+}3s)_2$ \\           
5 & 240.78 & 241.49 & 242.75 & 243.63 & \color{red} $240.770^b$ & & $(2s^22p_{-}2p_{+})_1$ & $(2s2p_{-}2p^2_{+})_1$ \\  
6 & 243.71 & 244.98 & 244.48 & 244.75 & \color{red} $243.760^b$ &  & $(2s^22p^2_{+})_2$ & $(2s2p_{-}2p^2_{+})_1$ \\
7$^*$ & 248.45 & 248.99 & 249.16 & 249.66 &   & 248.456 & $(2s^22p_{-}2p_{+})_2$ & $(2s2p^3_{+})_1$ \\ 
8 & 278.53 & 277.57 & 281.48 & 281.58 & \color{red} $278.699^b$ &  & $(2s^22p_{-}2p_{+})_2$ & $(2s2p_{-}2p^2_{+})_2$ \\ 
9 & 327.98 & 328.64 & 332.26  & 327.98 &  & 328.200 & $(2s^22p_{-}2p_{+})_1$ & $(2s2p^3_{+})_2$ \\          
10 & 332.51 & 333.52 & 335.51  & 331.50 & \color{red} $332.79^c$ &  & $(2s^22p^2_{+})_2$ & $(2s2p^3_{+})_2$  \\
\br
\end{tabular}
\begin{indented}
\item[] $^a$ Livermore EBIT experimental results \cite{Gu11}.
\item[] $^b$ Laser-produced plasma experimental results \cite{Valero72}.
\item[] $^c$ TJ-II stellarator plasma experimental results \cite{McCarthy16}.
\end{indented}
\end{table}

\clearpage

\subsection{Al$^{8+}$}
The atomic structure considered in our theoretical model for 
the B-like Al$^{8+}$ ion is listed in Table~\ref{table:Al8res}. 
All the mixing between these configurations are included in 
the calculations. 
\begin{table}[h!]
\caption{Description of the theoretical model used for the calculation of 
Al$^{8+}$ spectra. $I_{\mathrm{p}}$: Ionization Energy. 
$T_{\mathrm{max}}$: Temperature of maximum abundance. 
$N$: Number of levels considered in CRM, 
$N_{E1}$: Number of dipole-allowed radiative transitions. }
\begin{indented}
\item[]\begin{tabular}{ccccc}
\br
\multicolumn{1}{c|}{Ground Configuration} & 
\multicolumn{1}{c|}{$I_{\mathrm{p}}$ (eV)} & 
\multicolumn{1}{c|}{$T_{\mathrm{max}}$ (eV)} & 
\multicolumn{1}{c|}{$N$}  &
\multicolumn{1}{c}{$N_{E1}$}    \\
\cline{1-5}
\multicolumn{1}{c|}{$1s^{2}2s^{2}2p$ (B-like)}      &
\multicolumn{1}{c|}{330}      & 
\multicolumn{1}{c|}{75}      & 
\multicolumn{1}{c|}{281}   &
\multicolumn{1}{c}{11320}   \\ 
\cline{1-5}  
\mr
\multicolumn{1}{l}{Odd Configurations:} &
\multicolumn{4}{l}{  
$1s^{2}2s^{2}nl$ $(n=2,3,4,5 ~;~ l=1,3)$, }  \\
\multicolumn{1}{l}{} &
\multicolumn{4}{l}{ 
$1s^{2}2s2pnl$  ($n=3,4,5 ~;~ l=0,2,4)$ }   \\
\cline{1-5}
\multicolumn{1}{l}{Even Configurations:} &
\multicolumn{4}{l}{  
$1s^{2}2s^{2}nl$ $(n=3,4,5 ~;~ l=0,2,4)$, 
$1s^{2}2snl^{2}$ $(n=2,3 ~;~ l=0,1,2)$, }  \\
\multicolumn{1}{l}{} &
\multicolumn{4}{l}{ 
$1s^{2}2s4p^{2}$, 
$1s^{2}2s2pnl$ $(n=3,4,5 ~;~ l=1,3)$ }   \\
\br
\label{table:Al8res}
\end{tabular}
\end{indented}
\end{table}

\begin{center}
\begin{figure}[h!]
\centering\includegraphics[width=0.95\textwidth]{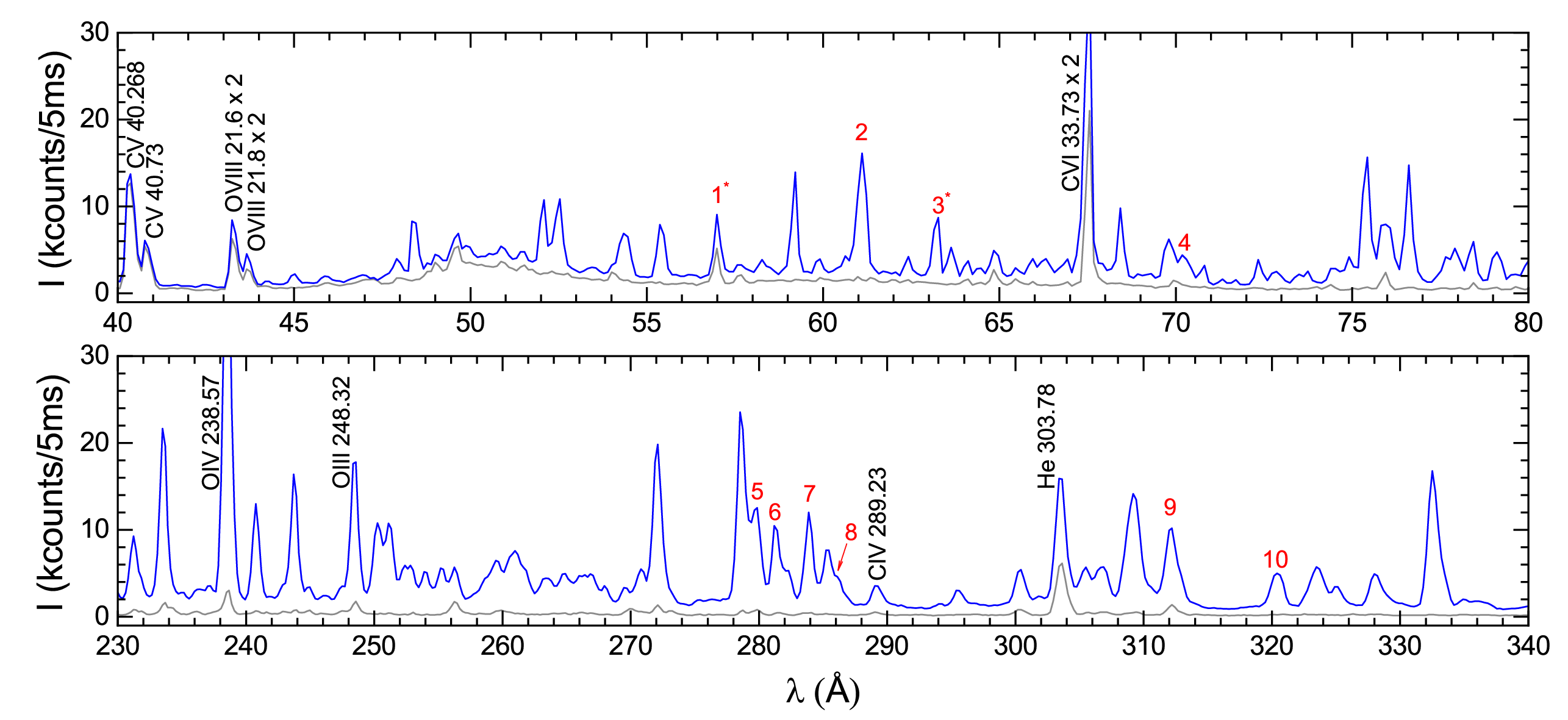}   
\caption{EAST spectrum and the lines identified as belonging to Al$^{8+}$.
Gray curve: spectrum before the Al burst. Blue curve: spectra at maximum Al emission. 
The numbers indicate the lines (Key) described in Table~\ref{table:Al8}.}
\label{fig:Al8}
\end{figure}                      
\end{center}

The higher transition probabilities are from radiative processes of the type $3d \rightarrow 2p$, in particular, the 
$2s3d^{2} \rightarrow 2s2p3d$ around the 57 \AA, and the $2s2p3d \rightarrow 2s2p^2$ in the 60-64 \AA~range, 
with radiative rate coefficients higher than $10^{12}$ s$^{-1}$.
The spectra lines identified for this ion are shown in Fig.~\ref{fig:Al8}.
With our CR model, we were able to determine that in the spectrum of this ion, line 2 in Table~\ref{table:Al8} ($3d \rightarrow 2p$) stands out for its intensity.  
In the 280-290 \AA~range, a group of high-intensity peaks is recognized both in the experimental and theoretical spectra (as $2p \rightarrow 2s$), labeled in the Table as 5 to 8, and the same transitions provide the high-intensity peaks 9 and 10. 
As mentioned before, line 3$^*$ overlaps with line 2$^*$ of Al$^{6+}$. Line 1$^*$ stands in the same wavelength as line 4$^*$ (Table~\ref{table:Al9}) of Al$^{9+}$.
Two experimental results from laser-produced plasmas \cite{Valero72} agree very well with our values (lines 6 and 9), 
but Valero and Goorvitch identified the lines as belonging to Al$^{4+}$ and Al$^{5+}$, respectively.
Our calculations can not reproduce any relevant line at these wavelengths for these ions. 
The wavelength reported by NIST for line 6 also cites the article \cite{Valero72} as a source, but we could not determine the origin of this data. 
Line 7 is marked with an asterisk since it fails at the same 
position as line 13 in Table \ref{table:Al6} (Al$^{6+}$).
The theoretical results given by the three codes show similar patterns.
However, it will be necessary to implement a more extended CR model to perform a thorough analysis that allows elucidation of the spectra in this region with greater confidence.

\begin{table}[h!]
\caption{EAST experimental and theoretical lines for Al$^{8+}$.
The column `Key' indicates the transitions labeled in Fig.~\ref{fig:Al8}.
We use the usual relativistic nomenclature $j_{\pm} \equiv l \pm \frac{1}{2}$. 
The values shown in red refer to lines assigned to other ions.     
Wavelengths are in \AA.}
\label{table:Al8}
\footnotesize
\begin{tabular}{lcccccccc}
\br
Key & Exp. & {\sc fac} & {\sc hullac} & AS & Previous Exp. & NIST & Lower & Upper \\
\mr
1$^*$ & 56.99 & 57.53 & 57.29 & 57.01 & $56.957^a$ & 56.945 & $(2s^22p_{+})_{3/2}$ &$(2s2p_{+}3p_{-})_{5/2}$ \\  
2 & 61.10 & 61.20 & 60.63 & 60.07 & $61.014^a$ & 61.078 & $(2s^22p_{+})_{3/2}$ & $(2s^23d_{+})_{5/2}$\\
3$^*$ & 63.27 & 63.69 & 63.56 & 62.78 & $63.627^a$ $63.509^b$ & 63.025 & $(2s2p_{-}2p_{+})_{5/2}$ & $(2s2p_{+}3d_{-})_{7/2}$ \\    
4 & 70.18 & 69.78 & 69.54 & 68.61 & $69.818^a$ & 69.716 & $(2s2p_{-}2p_{+})_{5/2}$ & $(2s2p_{+}3s)_{3/2}$ \\ 
5 & 279.86 & 283.22 & 281.55 & 283.89 &  & 280.114 & $(2s^22p_{-})_{1/2}$ & $(2s2p^2_{+})_{3/2}$ \\  
6 & 281.42 & 285.62 & 283.99 & 286.70 & \color{red} $281.433^b$ & 282.407 & $(2s^22p_{-})_{1/2}$ & $(2s2p_{-}2p_{+})_{1/2}$ \\ 
7$^*$ & 283.88 & 287.28	& 285.53 & 288.45 & $284.04^c$ & 284.015 & $(2s^22p_{+})_{3/2}$ & $(2s2p^2_{+})_{3/2}$ \\ 
8 & 286.13 & 289.76	& 288.03 & 291.35 &  $286.364^c$& 286.364 & $(2s^22p_{+})_{3/2}$ & $(2s2p_{-}2p_{+})_{1/2}$ \\
9 & 312.21 & 313.48 & 313.46 & 312.11 & \color{red} $312.241^b$ &    &  $(2s^22p_{-})_{1/2}$ & $(2s2p^2_{+})_{1/2}$ \\ 
10 & 320.41 & 318.47 & 318.39 & 317.63 &    & 321.027 &  $(2s^22p_{+})_{3/2}$ & $(2s2p^2_{+})_{1/2}$ \\
\br
\end{tabular}
\begin{indented}
\item[] $^a$ Livermore EBIT experimental results \cite{Gu11}.
\item[] $^b$ Laser-produced plasma experimental results \cite{Valero72}.
\item[] $^c$ TJ-II stellarator plasma experimental results \cite{McCarthy16}.
\end{indented}
\end{table}
\clearpage
\subsection{Al$^{9+}$}

The Be-like ion Al$^{9+}$ has a simple atomic structure in which the $1s^22s^2$ is the ground level. 
The ionization energy is 398.7 eV, and the temperature of maximum abundance is around 90 eV.
To identify the lines, we performed a CR model calculation, 
in which we included the configurations listed in Table~\ref{table:Al9res}, 
taking into acount all the mixing between these configurations. 
\begin{table}[h!]
\caption{Description of the theoretical model used for the calculation of 
Al$^{9+}$ spectra. $I_{\mathrm{p}}$: Ionization Energy. 
$T_{\mathrm{max}}$: Temperature of maximum abundance. 
$N$: Number of levels considered in CRM, 
$N_{E1}$: Number of dipole-allowed radiative transitions. }
\begin{indented}
\item[]\begin{tabular}{ccccc}
\br
\multicolumn{1}{c|}{Ground Configuration} & 
\multicolumn{1}{c|}{$I_{\mathrm{p}}$ (eV)} & 
\multicolumn{1}{c|}{$T_{\mathrm{max}}$ (eV)} & 
\multicolumn{1}{c|}{$N$}  &
\multicolumn{1}{c}{$N_{E1}$}    \\
\cline{1-5}
\multicolumn{1}{c|}{$1s^{2}2s^{2}$ (Be-like)}      &
\multicolumn{1}{c|}{399}      & 
\multicolumn{1}{c|}{90}      & 
\multicolumn{1}{c|}{94}   &
\multicolumn{1}{c}{1237}   \\ 
\cline{1-5}  
\mr
\multicolumn{1}{l}{Even Configurations:} &
\multicolumn{4}{l}{  
$1s^{2}2s^{2}$, $1s^{2}2snl$  ($n=3,4,5 ~;~ l=0,2,4)$, }  \\
\multicolumn{1}{l}{} &
\multicolumn{4}{l}{ 
$1s^{2}2p^{2}$, $1s^{2}3s^{2}$, $1s^{2}3s^{2}$, $1s^{2}4p^{2}$, 
$1s^{2}4d^{2}$, $1s^{2}5s^{2}$}   \\
\cline{1-5}
\multicolumn{1}{l}{Odd Configurations:} &
\multicolumn{4}{l}{  
$1s^{2}2snl$ $(n=2,3,4,5 ~;~ l=1,3)$, }\\
\multicolumn{1}{l}{} &
\multicolumn{4}{l}{ 
$1s^{2}2p3s$, $1s^{2}2p3d$ }   \\
\br
\label{table:Al9res}
\end{tabular}
\end{indented}
\end{table}

\begin{center}
\begin{figure}[h!]
\centering\includegraphics[width=0.95\textwidth]{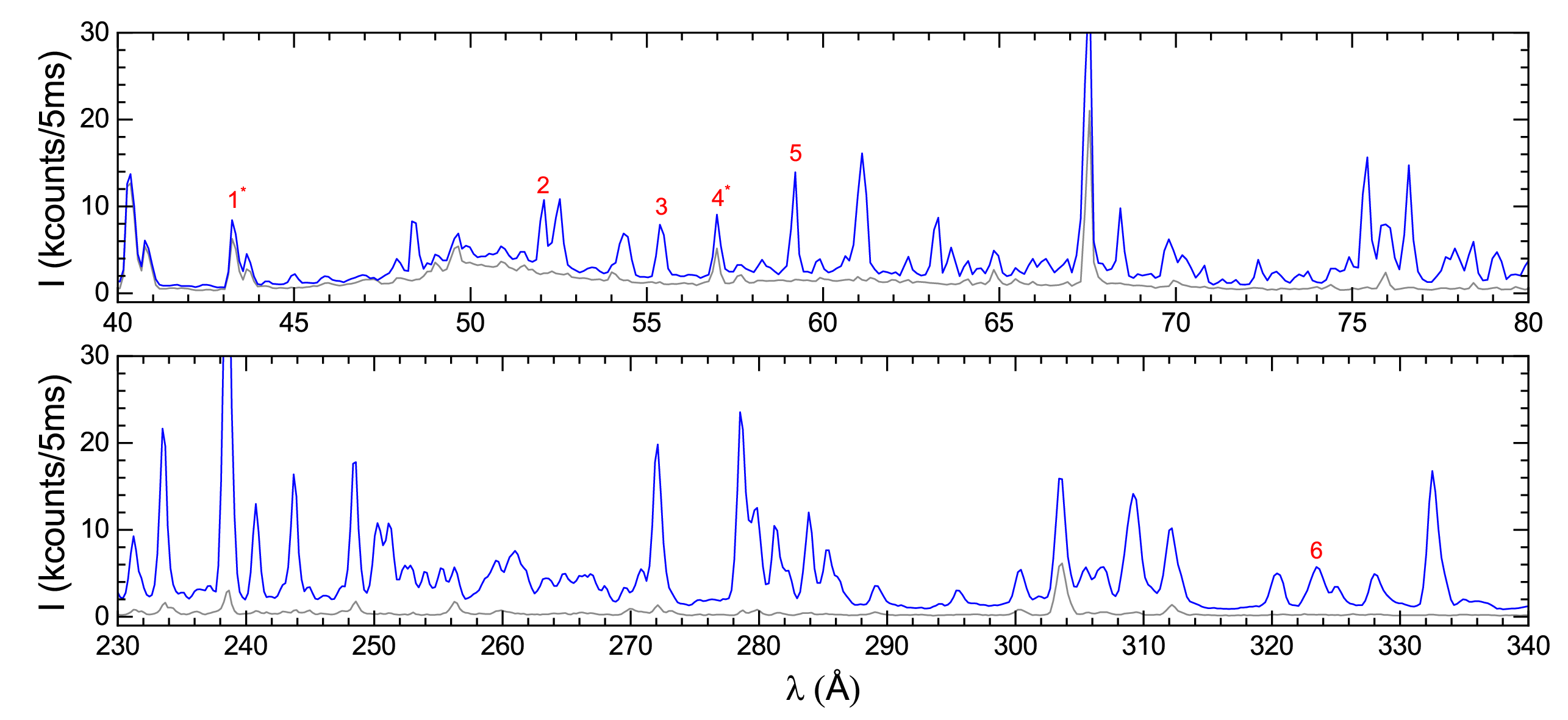}   
\caption{EAST spectrum and the lines identified as belonging to Al$^{9+}$.
Gray curve: spectrum before the Al burst. Blue curve: spectra at maximum Al emission. 
The numbers indicate the lines (Key) described in Table~\ref{table:Al9}.}
\label{fig:Al9}
\end{figure}                      
\end{center}

There are fewer than 30 transitions with radiative rate coefficients greater than  $10^{12}$ s$^{-1}$, and they are mainly $2snl \rightarrow 2s2p$ (between 38 \AA~to 45 \AA) and $2pnl \rightarrow 2p^2$ (between 56 \AA~to 60 \AA). 
Only two transitions to the ground level have high Einstein coefficients, these are the lines labeled 
2 and 6 in Table~\ref{table:Al9}, $3p \rightarrow 2s$ and $2p \rightarrow 2s$, respectively.
The spectral lines corresponding to this ion are marked in Fig.~\ref{fig:Al9}.
Our calculations show that only the six lines listed in Table~\ref{table:Al9} are produced by radiative emission of the Al$^{9+}$ ion.
The first line listed (1$^*$ at 43.24 \AA) overlaps with the second-order OVII line ($2 \times 21.6$ \AA), 
and line 4$^*$ overlaps with line 1$^*$ of Table~\ref{table:Al8} corresponding to Al$^{8+}$. 
The line marked as 6 ($2s2p \rightarrow 2s^2$, with wavelengths close to 324 \AA) deserves special attention. 
On the one hand, the three calculations have turned out with quite similar wavelengths and in excellent agreement with the experimental value of 323.47 A (errors of 0.15\%, 0.09\%, and 0.01\% for {\sc fac}, {\sc hullac}, and {\sc as}, respectively). However, the resulting synthetic spectra of our CR model predict a sharp line with enormous intensity, much larger than the other spectral lines emitted by this ion.  
That is not reflected in the experimental spectrum of EAST. Moreover, as already mentioned in Section \ref{subsec:Al7}, both the stellarator plasma experiments of McCarthy {\it et al.} \cite{McCarthy16} as well as the NIST database report that this transition appears at 333 \AA. 
In our experiments, a very strong spectral line arises at this wavelength, but our calculations attribute it to Al$^{7+}$, as reported in Table \ref{table:Al7}.  
We do not have any theoretical explanation for a possible quenching of 
the ($2s2p \rightarrow 2s^2$) line.
If our wavelength calculations become correct, we must perform a deep theoretical analysis to understand the physical mechanisms that rule the radiation at this spectrum region.
On the contrary, if our three calculations were to mispredict the wavelengths by about 10 \AA, we would need to delve thoroughly into the atomic structure of this ion to understand the causes of such a significant error. We have tested various configuration interactions, but we did not manage to approach the value of 332 \AA. 

\begin{table}[h!]
\caption{EAST experimental and theoretical lines for Al$^{9+}$.
The column `Key' indicates the transitions labeled in Fig.~\ref{fig:Al9}. 
Wavelengths are in \AA.}
\label{table:Al9}
\footnotesize
\begin{indented}
\item[]\begin{tabular}{lcccccccc}
\br
Key & Exp. & {\sc fac} & {\sc hullac} & AS & Previous Exp. & NIST & Lower & Upper \\
\mr               
1$^*$ & 43.24 & 42.43 & 42.38 & 42.28 &  &  & $(2s2p_{+})_2$ & $(2s4d_{+})_3$ \\ 
2 & 52.09 & 52.89 & 51.93 & 51.80 & $51.975^a$ & 51.979 & $2s^2$ &$(2s3p_{-})_1$ \\                       
3 & 55.38 & 55.37 & 55.33 & 55.18 & $55.318^a$ & 55.376 & $(2s2p_{+})_2$ & $(2s3d_{+})_3$ \\     
4$^*$ & 56.99 & 56.96 & 57.13 & 56.82 & $56.957^a$ & 56.945 & $(2p^2_{+})_2$ &$(2p_{+}3d_{+})_3$ \\  
5 & 59.21 & 59.35 & 59.19 & 59.02 & $59.094^a$ $59.107^b$ & 59.298 & $(2s2p_{+})_1$ & $(2s3d_{+})_2$ \\     
6 & 323.47 & 322.94 & 323.78 & 323.51 &   &  & $2s^2$  & $(2s2p_{+})_1$   \\
\br                  
\end{tabular}
\item[] $^a$ Livermore EBIT experimental results \cite{Gu11}.
\item[] $^b$ Laser-produced plasma experimental results \cite{Valero72}.
\end{indented}
\end{table}

\clearpage
\subsection{Al$^{10+}$}

\begin{center}
\begin{figure}[h!]
\centering\includegraphics[width=0.95\textwidth]{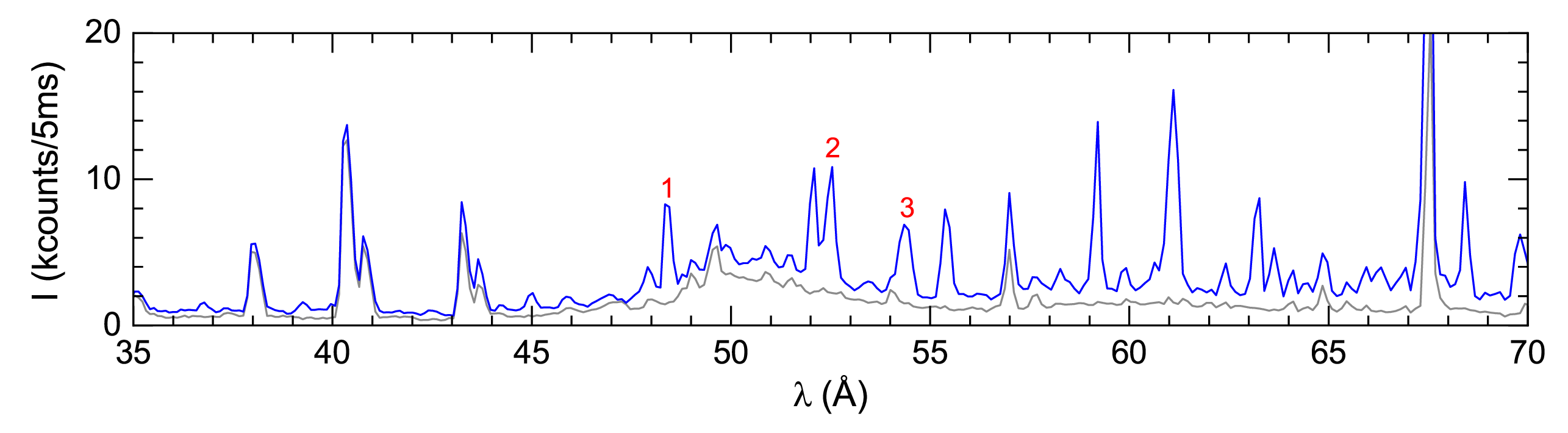}   
\caption{EAST spectrum and the lines identified as belonging to Al$^{10+}$.
Gray curve: spectrum before the Al burst. Blue curve: spectra at maximum Al emission. 
The numbers indicate the lines (Key) described in Table~\ref{table:Al5}.}
\label{fig:Al10}
\end{figure}                      
\end{center}

This ion, isoelectronic to the Li (ground $1s^22s$), has a straightforward  atomic structure and a few radiative lines, as is shown in Fig.~\ref{fig:Al10}.
The ionization energy is 442 eV, and the maximum abundance is about 116 eV.
We implemented a simple CR model, with only 24 levels belonging to the 
configurations listed in Table~\ref{table:Al10res}, 
including all the mixing between them, 
resulting all the dipole-allowed radiative transitions. 
\begin{table}[h!]
\caption{Description of the theoretical model used for the calculation of 
Al$^{10+}$ spectra. $I_{\mathrm{p}}$: Ionization Energy. 
$T_{\mathrm{max}}$: Temperature of maximum abundance. 
$N$: Number of levels considered in CRM, 
$N_{E1}$: Number of dipole-allowed radiative transitions. }
\begin{indented}
\item[]\begin{tabular}{ccccc}
\br
\multicolumn{1}{c|}{Ground Configuration} & 
\multicolumn{1}{c|}{$I_{\mathrm{p}}$ (eV)} & 
\multicolumn{1}{c|}{$T_{\mathrm{max}}$ (eV)} & 
\multicolumn{1}{c|}{$N$}  &
\multicolumn{1}{c}{$N_{E1}$}    \\
\cline{1-5}
\multicolumn{1}{c|}{$1s^{2}2s$ (Li-like)}      &
\multicolumn{1}{c|}{442}      & 
\multicolumn{1}{c|}{116}      & 
\multicolumn{1}{c|}{24}   &
\multicolumn{1}{c}{92}   \\ 
\cline{1-5}  
\mr
\multicolumn{1}{l}{Even Configurations:} &
\multicolumn{4}{l}{  
$1s^{2}nl$  ($n=2,3,4,5 ~;~ l=0,2,4)$ }  \\
\cline{1-5}
\multicolumn{1}{l}{Odd Configurations:} &
\multicolumn{4}{l}{  
$1s^{2}nl$ $(n=2,3,4,5 ~;~ l=1,3)$ }\\
\br
\label{table:Al10res}
\end{tabular}
\end{indented}
\end{table}

Table~\ref{table:Al10} lists the three dominant lines that appear highly in the spectra.
The agreement between the three different calculations and the experimental results is excellent. 
In particular, {\sc hullac} results are closer than 0.2\% from the measured values.

\begin{table}[h!]
\caption{EAST experimental and theoretical lines for Al$^{10+}$.
The column `Key' indicates the transitions labeled in Fig.~\ref{fig:Al10}. 
Wavelengths are in \AA.}
\label{table:Al10}
\begin{indented}
\item[]\begin{tabular}{ccccccccc}
\br
Key & Exp. & {\sc fac} & {\sc hullac} & AS & Previous Exp. & NIST & Lower & Upper \\
\mr
1 & 48.34 & 48.29 & 48.35 & 48.27 & $48.330^a$ & 48.297 & $2s$ & $3p_{-}$ \\     
2 & 52.54 & 52.44 & 52.49 & 52.45 & $52.438^a$ & 52.446 & $2p_{+}$ & $3d_{+}$ \\     
3 & 54.35 & 54.41 & 54.43 & 54.41 & $54.219^a$ & 54.388 & $2p_{+}$ & $3s$ \\
\br
\end{tabular}
\item[] $^a$ Livermore EBIT experimental results \cite{Gu11}.
\end{indented}
\end{table}

\clearpage
\subsection{Al$^{11+}$ and Al$^{12+}$}

The Al$^{11+}$ ion is isoelectronic to He (ground configuration $1s^2$). 
Therefore, it has relatively considerable ionization energy (2086 eV) and is the dominant ion along a broad range of temperatures, roughly 70 to 1000 eV. 
The structure included in the CR model is listed in Table~\ref{table:Al11res},
taking into account all the mixing between these configurations, 
obtaining very accurate results, as shown in Table~\ref{table:Al11}.
\begin{table}[h!]
\caption{Description of the theoretical model used for the calculation of 
Al$^{11+}$ spectra. $I_{\mathrm{p}}$: Ionization Energy. 
$T_{\mathrm{max}}$: Temperature of maximum abundance. 
$N$: Number of levels considered in CRM, 
$N_{E1}$: Number of dipole-allowed radiative transitions. }
\begin{indented}
\item[]\begin{tabular}{ccccc}
\br
\multicolumn{1}{c|}{Ground Configuration} & 
\multicolumn{1}{c|}{$I_{\mathrm{p}}$ (eV)} & 
\multicolumn{1}{c|}{$T_{\mathrm{max}}$ (eV)} & 
\multicolumn{1}{c|}{$N$}  &
\multicolumn{1}{c}{$N_{E1}$}    \\
\cline{1-5}
\multicolumn{1}{c|}{$1s^{2}$ (He-like)}      &
\multicolumn{1}{c|}{2086}      & 
\multicolumn{1}{c|}{350}      & 
\multicolumn{1}{c|}{49}   &
\multicolumn{1}{c}{336}   \\ 
\cline{1-5}  
\mr
\multicolumn{1}{l}{Even Configurations:} &
\multicolumn{4}{l}{  
$1s^{2}$, $1snl$  ($n=2,3,4,5 ~;~ l=0,2,4)$ }  \\
\cline{1-5}
\multicolumn{1}{l}{Odd Configurations:} &
\multicolumn{4}{l}{  
$1snl$ $(n=2,3,4,5 ~;~ l=1,3)$ }\\
\br
\label{table:Al11res}
\end{tabular}
\end{indented}
\end{table}

In the EAST experiment, only the $K\alpha$ lines have been detected at the low edge of the wavelengths that our instruments can measure.
The $1s3p \rightarrow 1s^2$ and $1s4p \rightarrow 1s^2$ lines should appear at 6.6 \AA~and 6.3 \AA, but these can not be distinguished from the background.

Concerning the H-like ion Al$^{12+}$, 
For the theoretical model, we included the configurations listed 
in Table~\ref{table:Al12res}, allowing us to accurately calculate 
the wavelengths. 
\begin{table}[h!]
\caption{Description of the theoretical model used for the calculation of 
Al$^{12+}$ spectra. $I_{\mathrm{p}}$: Ionization Energy. 
$T_{\mathrm{max}}$: Temperature of maximum abundance. 
$N$: Number of levels considered in CRM, 
$N_{E1}$: Number of dipole-allowed radiative transitions. }
\begin{indented}
\item[]\begin{tabular}{ccccc}
\br
\multicolumn{1}{c|}{Ground Configuration} & 
\multicolumn{1}{c|}{$I_{\mathrm{p}}$ (eV)} & 
\multicolumn{1}{c|}{$T_{\mathrm{max}}$ (eV)} & 
\multicolumn{1}{c|}{$N$}  &
\multicolumn{1}{c}{$N_{E1}$}    \\
\cline{1-5}
\multicolumn{1}{c|}{$1s$ (H-like)}      &
\multicolumn{1}{c|}{1039}      & 
\multicolumn{1}{c|}{2300}      & 
\multicolumn{1}{c|}{25}   &
\multicolumn{1}{c}{100}   \\ 
\cline{1-5}  
\mr
\multicolumn{1}{l}{Even Configurations:} &
\multicolumn{4}{l}{  
$1s$, $nl$  ($n=2,3,4,5 ~;~ l=0,2,4)$ }  \\
\cline{1-5}
\multicolumn{1}{l}{Odd Configurations:} &
\multicolumn{4}{l}{  
$nl$ $(n=2,3,4,5 ~;~ l=1,3)$ }\\
\br
\label{table:Al12res}
\end{tabular}
\end{indented}
\end{table}

For this ion, only the $K\alpha_1$ and $K\alpha_2$ lines are visible 
in the EAST spectra, but these are not fully resolved in our measurements, 
so they appear as a unique peak at 7.18 \AA.
The lines indicated in Table~\ref{table:Al11}, have also been observed in laser-produced plasma experiments  \cite{Boiko77,Couturaud77,Boiko78,Shlyaptsev16,Channprit17}, in excellent agreement with our values.

\begin{table}[h!]
\caption{EAST experimental and theoretical lines for Al$^{11+}$ and Al$^{12+}$.
Wavelengths are in \AA.}
\label{table:Al11}
\begin{indented}
\item[]\begin{tabular}{ccccccccc}
\br
Ion  & Exp. & {\sc fac} & {\sc hullac} & AS & Previous Exp. & NIST & Lower & Upper \\
\mr
Al$^{12+}$ & ~ & 7.171 &   7.171   &   7.169   & $7.1703^a$  $7.17^{b,c}$  &   7.171   & $1s$ &	$2p_{+}$ \\ 
           & 7.18  & 7.176 &   7.176   &   7.174   & $7.1759^a$ &   7.177   & $1s$ &	$2p_{-}$ \\ 
\mr
Al$^{11+}$ & ~ & 7.759 &   7.754   &   7.769  & $7.75^d$ &   7.758   & $1s^2$ &	$(1s2p_{+})_1$ \\ 
           & 7.80 & 7.812 &   7.809   &  7.820   & $7.80^d$  $7.81^e$ &    7.807   & $1s^2$ &	$(1s2p_{-})_1$ \\
\br
\end{tabular}
\item[] $^a$ Laser produced plasma experimental results \cite{Boiko77}.
\item[] $^b$ Laser produced plasma experimental results \cite{Couturaud77}.
\item[] $^c$ Laser produced plasma experimental results \cite{Channprit17}.		
\item[] $^d$ Laser produced plasma experimental results \cite{Boiko78}.
\item[] $^e$ Laser produced plasma experimental results \cite{Shlyaptsev16}.
\end{indented}
\end{table}

\clearpage

\section{Summary}
\label{sec:summary}

Extreme ultraviolet (EUV) spectra emitted from aluminum ions in the 5–340 \AA~wavelength range were observed in Experimental Advanced Superconducting Tokamak (EAST) discharges. 
We use three different atomic computational codes, {\sc fac}, {\sc hullac}, and {\sc autostructure}, to identify the lines observed in our measurements. 
The spectral lines have been identified by means of a collisional-radiative model that treats each ion independently. For each ion, the excited states are assumed to be in equilibrium with the ground state through electron-impact excitation and de-excitation processes, and radiative transitions (dipole-allowed). This model is complemented by a fractional ion abundance calculation (coronal approximation), which allows each temperature to be associated with a global factor by which the line strengths of each ion are multiplied. 
Overall, we found that the three codes can reproduce the experimental wavelengths with very good accuracy. 
Especially if consider that the calculations have been made without any additional adjustments since our objective has been to evaluate the capabilities of these codes.
The performance of the {\sc hullac} code is particularly remarkable, which in many cases reproduces the experimental values with errors less than 0.3\%.
\section{Acknowledgments}

This work was supported by the National MCF Energy R~\&~D Program (Nos. 2022YFE03180400, 2019YFE030403, 2018YFE0311100), National Natural Science Foundation of China (Grant Nos. 12322512, 11905146, 11975273) and Chinese Academy of Sciences President’s International Fellowship Initiative (PIFI) (No. 2020VMA0001).
DM acknowledges partial support from CONICET by Project No.
PIP11220200102421CO, and the ANPCyT by Project No. PICT-
2020-SERIE A-01931 in Argentina, and the Alliance of International Science Organizations (ANSO) Visiting Fellowship (ANSO-VF-2021-03), in China.


\end{document}